\documentclass[]{elsart}
\usepackage{psfig}
\usepackage{amssymb}

\begin{document}
 \begin{frontmatter}
  \title{On the recent star formation history of the Milky Way disk}
  \author{R. de la Fuente Marcos}
   \address{Suffolk University Madrid Campus,
            C/ Vi\~na, 3. E-28003 Madrid, Spain \\
            raul@galaxy.suffolk.es}
  \author{C. de la Fuente Marcos}
   \address{Universidad Complutense de Madrid, E-28040 Madrid, Spain \\ 
            nbplanet@fis.ucm.es}


  \begin{abstract}
     We have derived the star formation history of the Milky Way disk
     over the last 2 Gyr with a time resolution of 0.05 Gyr from the
     age distribution diagram of a large sample of open clusters
     comprising more than 580 objects. By interpreting the age 
     distribution diagram using numerical results from an extensive 
     library of $N$-body calculations carried out during the last ten 
     years, we reconstruct the recent star formation history of the 
     Milky Way disk. Under the assumption that the disk has never been 
     polluted by any extragalactic stellar populations, our analysis 
     suggests that superimposed on a relatively small level of constant 
     star formation activity mainly in small-$N$ star clusters, the star 
     formation rate has experienced at least 5 episodes of enhanced star 
     formation lasting about 0.2 Gyr with production of larger clusters. 
     This cyclic behavior shows a period of 0.4$\pm$0.1 Gyr and could be 
     the result of density waves and/or interactions with satellite 
     galaxies. On the other hand, the star formation rate history from a 
     volume-limited sample of open clusters in the solar neighbourhood 
     appears to be consistent with the overall star formation history 
     obtained from the entire sample. Pure continuous star formation both 
     in the solar neighbourhood and the entire Galactic disk is strongly 
     ruled out. Our results also indicate that, in the Milky Way disk, 
     about 90\% of open clusters are born with $N \leq 150$ and the slope 
     in the power-law frequency distribution of their masses is about -2.7 
     when quiescent star formation takes place. If the above results are 
     re-interpreted taking into consideration accretion events onto the 
     Milky Way, it is found that a fraction of the unusually high number 
     of open clusters with ages older than 0.6 Gyr may have been formed 
     in disrupted satellites. Problems arising from the selection effects 
     and the age errors in the open cluster sample used are discussed in 
     detail. 
  \end{abstract}

 \end{frontmatter}

 \section{Introduction} \label{intro}
    Throughout the Milky Way's history, a non-negligible fraction of star
    formation has apparently occurred in starburst-like events (see
    Majewski, 1993 or Freeman \& Bland-Hawthorn, 2002, for a review). 
    On the other hand, observations suggest that present-day star 
    formation in the disk of our Galaxy takes place in stellar 
    groupings rather than in isolation. These stellar aggregates form 
    from molecular clouds, and in the disk of the Milky Way they appear
    in two types: bound and unbound. Unbound, short lived stellar 
    groupings are called associations; bound, long lived stellar groups 
    are known as open clusters. Open clusters can also be formed out of 
    the remains of rich stellar associations (Kroupa et al., 2001). 
    As most of the field stars appear to have been formed in the 
    so-called clustered mode (i.e., in clusters or associations), not 
    in the dispersed mode (i.e., in isolation), it is a natural choice
    to consider these stellar clumpings as the {\it de facto} units of 
    star formation in the disk of our Galaxy (Clarke et al., 2000). The
    idea of open clusters being fundamental units of star formation in
    the Galactic disk is, however, controversial (Meyer et al., 2000).
    It has been argued that bound open clusters cannot contribute 
    significantly to the field star population of the Galactic disk
    because they are rare and long lived (e.g. Roberts, 1957). In 
    contrast, most young embedded clusters are thought to evolve into
    unbound stellar associations, which comprise the majority of stars
    that populate the Milky Way disk (Lada \& Lada, 1991). Nevertheless,
    if star clusters are the elementary units of star formation they 
    can, in principle, be used to derive the star formation history, 
    recent and old. Unfortunately, stellar associations evolve and 
    dissolve in a time-scale of $\sim$ 50 Myr (Brown, 2002), therefore 
    they cannot be used to study the star formation history of the 
    Galactic disk. On the other hand, open clusters are comparatively 
    long lived objects that may serve as excellent probes into the 
    structure and evolution of the Galactic disk.

    The problem of deriving the star formation history of the Milky
    Way has been considered by a number of authors using different
    techniques (Twarog, 1980; Scalo, 1987; Barry, 1988; Gomez et al., 
    1990; Marsakov et al., 1990; Noh \& Scalo, 1990; Meusinger, 1991; 
    Soderblom et al., 1991; Micela et al., 1993; D\'{\i}az-Pinto et 
    al., 1994; Rocha-Pinto \& Maciel, 1997; Chereul et al., 1998; 
    Isern et al., 1999; Lachaume et al., 1999; Rocha-Pinto et al., 
    2000a; Hernandez et al., 2000a; Bertelli \& Nasi, 2001; Gizis et 
    al., 2002; Just, 2002, 2003). The majority of these studies suggest 
    that the disk of the Milky Way Galaxy has not experienced a smooth 
    and constant star formation history but a bursty one with several 
    episodes of enhanced star formation. Although most of these papers 
    are restricted to the study of the star formation history in the 
    solar neighbourhood, the star sample birth sites are in fact 
    distributed over a larger range of distances because of orbital 
    diffusion, and so they can provide an estimate of the global star 
    formation rate and their conclusions can be extrapolated to the 
    entire Milky Way disk.  

    In this paper we revisit the topic of the recent star formation
    history of the Milky Way disk by using data from the {\it Open
    Cluster Database} (Mermilliod, 2003), the {\it New Catalogue of 
    Optically Visible Open Clusters and Candidates} (Dias et al., 2003a, b)
    as well as {\it Hipparcos} data (ESA, 1997) to construct the open 
    cluster age distribution. The age distribution diagram is then 
    interpreted using numerical results from an extensive library of 
    $N$-body calculations carried out during the last ten years. This 
    method permits the reconstruction of the star formation history 
    of the Milky Way disk over the last 2 Gyr with a time resolution 
    of 0.05 Gyr using very few a priori assumptions. Our main objective 
    is to understand the recent star formation history of the entire 
    Galactic disk although only the solar circle ($\pm$ 1.75 kpc) can
    be studied due to the incompleteness of the available open cluster
    sample (see Section 2).

    This paper is organized as follows: in Section 2, we present the 
    properties of the Galactic open cluster sample. The raw histogram
    of open cluster number vs. age is also included in this Section as
    well as a discussion of the potential limitations of our approach.
    In Section 3, we summarize some relevant results on the evolution 
    and dissolution of realistic $N$-body star cluster models. This 
    Section also includes a comparison between published open cluster 
    ages and number of stars with the results of the numerical models. 
    In Section 4, we present our method as well as results for the 
    Galactic disk. A closer view to the last 0.2 Gyr and the derived 
    open cluster initial mass function are considered in Section 5. The 
    star formation history is presented and discussed in Section 6; 
    corrections are also discussed here. A detailed comparison with 
    results from other authors is carried out in Section 7. An alternative 
    interpretation of our results in the context of dynamical merger 
    histories is included in Section 8. Open questions and conclusions 
    are summarized in Section 9.

 \section{An open cluster sample} \label{sample}
\begin{figure*}
        \centerline{\hbox{
        \psfig{figure=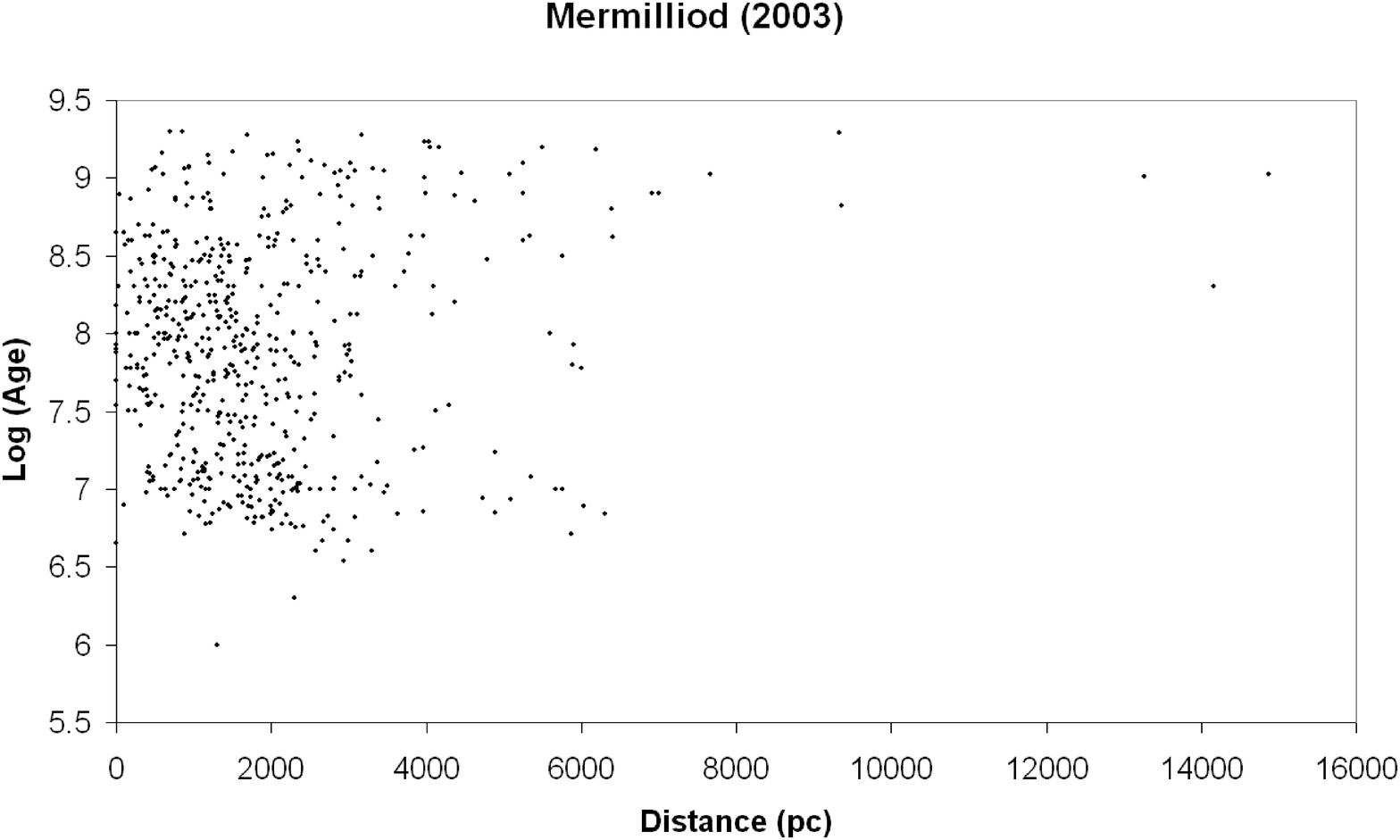,height=12cm,width=14cm,angle=0}
        }}
   \caption[]{Age-distance diagram of the open cluster sample from 
              the latest update (October 2003) of WEBDA (Mermilliod, 
              2003).  Only clusters with age $\leq$ 2 Gyr (581 objects)
              are plotted.  The sample appears to be rather incomplete 
              for open clusters more distant than 3.5 kpc.  
             }
      \label{mermilliod}
\end{figure*}

\begin{figure*}
        \centerline{\hbox{
        \psfig{figure=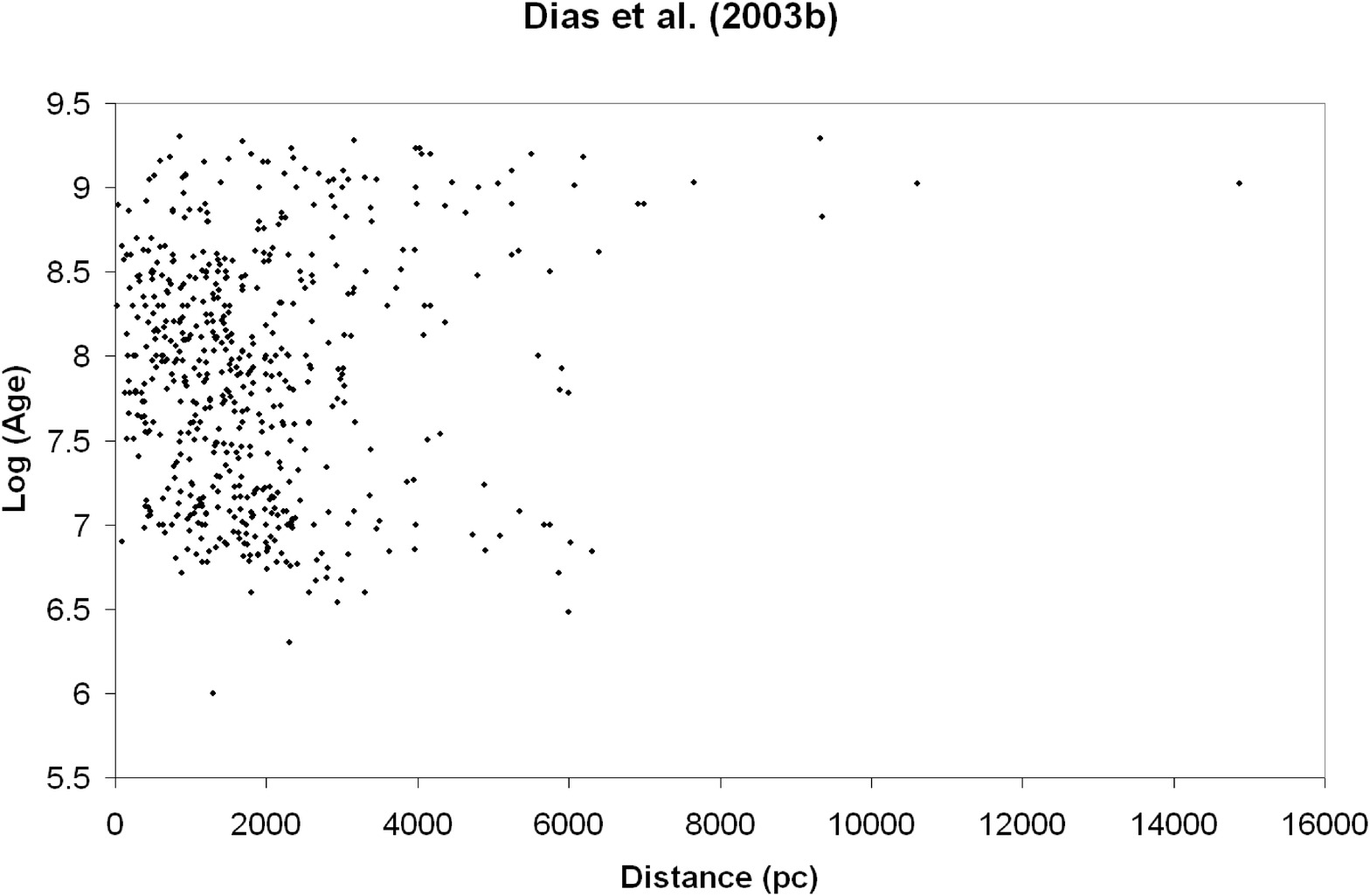,height=12cm,width=14cm,angle=0}
        }}
   \caption[]{Age-distance diagram of the open cluster sample from 
              the latest version of the {\it New Catalogue of 
              Optically Visible Open Clusters and Candidates} (Dias 
              et al., 2003b) for clusters of age $\leq$ 2 Gyr (569
              objects). There is a strong selection effect as
              the number of open clusters more distant than 3.5 kpc
              is small in our sample.  
             }
      \label{dias}
\end{figure*}

\begin{figure*}
        \centerline{\hbox{
        \psfig{figure=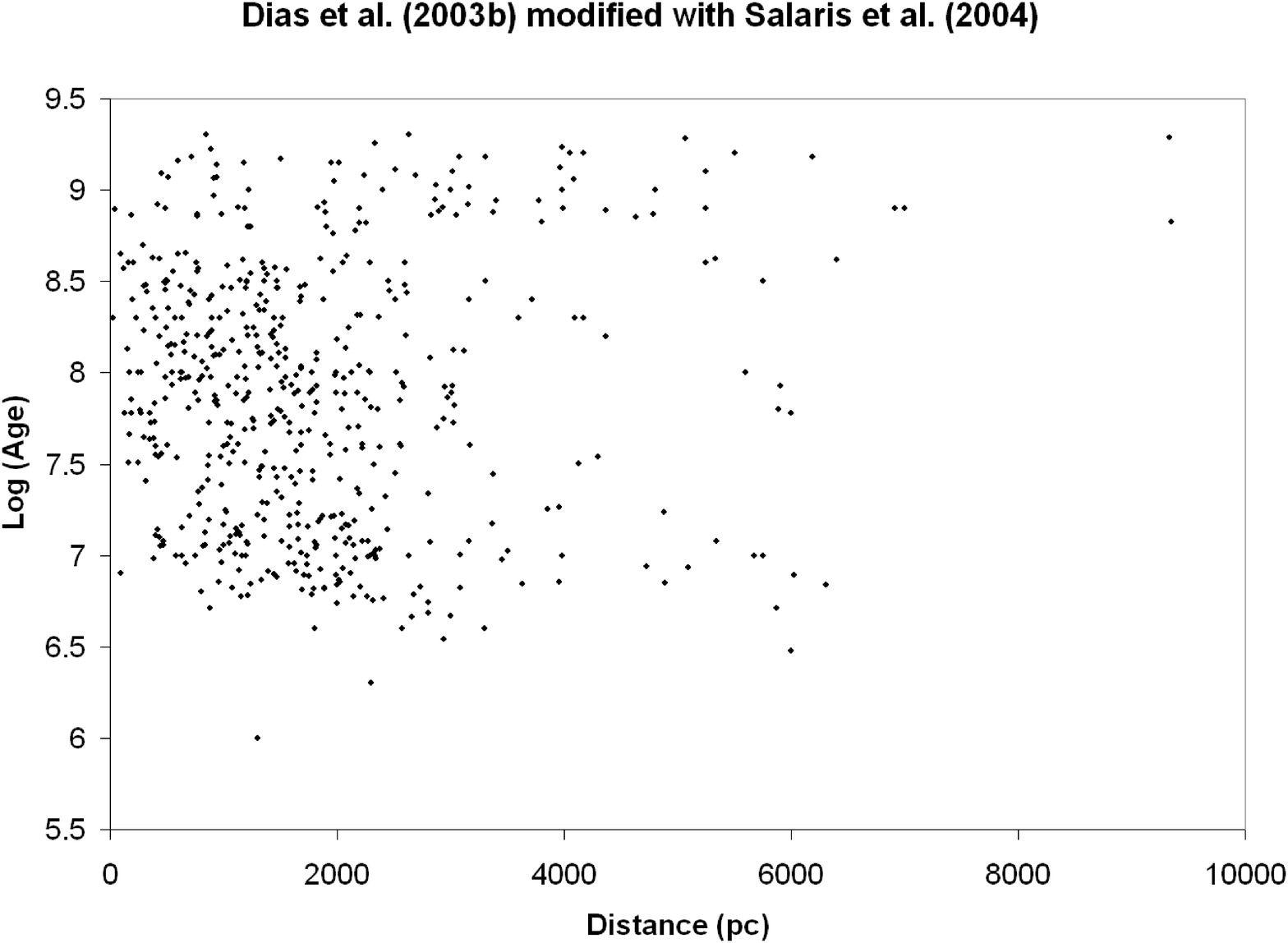,height=12cm,width=14cm,angle=0}
        }}
   \caption[]{Age-distance diagram of the open cluster sample 
              (559 objects) in Fig. \ref{dias} 
              modified with new age determinations by Salaris et 
              al. (2004). This sample includes less objects and
              covers a different distance range because some of
              the new ages are older ($>$ 2 Gyr) than those in
              the two previous samples.   
             }
      \label{salaris}
\end{figure*}

    For many years, the Lyng\aa \ catalogue (1985, 1987) has been the 
    classical reference source for open cluster data. An exhaustive
    analysis of the properties of the Milky Way open cluster system 
    using this catalogue was carried out by Janes et al. (1988). 
    Unfortunately, the Lyng\aa \ catalogue is not the most updated 
    source of data in this field. The {\it Open Cluster Database} 
    (WEBDA) created and maintained by J.-C. Mermilliod (1988, 1992a, 
    1992b, 1993, 1995, 1996, 2003) includes all the data already covered 
    in Lyng\aa's catalogue and many more. The latest update of the Open 
    Cluster Database (October 2003) includes 1731 open clusters with 
    ages for 616 objects (36\%). In this database we have found 581
    clusters with age $\leq$ 2 Gyr. Slightly less complete is the 
    {\it New Catalogue of Optically Visible Open Clusters and Candidates} 
    (NCOVOCC). The first version of this catalogue was published in Dias 
    et al. (2002). The October 2003 version (Dias et al., 2003b), includes 
    1637 objects, 603 (37\%) with published ages. This new catalogue 
    updates the previous catalogues of Lyng\aa \ (1987) and of Mermilliod 
    (1995) included in the WEBDA database. There are however non-negligible
    differences between the two databases as we can see from Figs.
    \ref{mermilliod} and \ref{dias}. NCOVOCC includes 569 clusters of 
    age $\leq$ 2 Gyr. On the other hand, Salaris et al. (2004) have recently 
    re-determined ages for a sample of 71 of the oldest open clusters (all 
    of them older than 0.67 Gyr) using a morphological age indicator, 
    metallicities, and {\it Hipparcos} parallaxes. In general, the new ages 
    are older than those from WEBDA and NCOVOCC. In order to study the impact 
    of different age determination techniques on our results we have modified 
    the NCOVOCC data with the new age determinations from Salaris et al. 
    (2004). The new subsample consists of 559 open clusters younger than 
    2 Gyr.  The age-distance diagram for this new sample appears in Fig. 
    \ref{salaris}. 
   
%
%
%
%
%
%
%
%
%
%
%
%

\begin{figure*}
        \centerline{\hbox{
        \psfig{figure=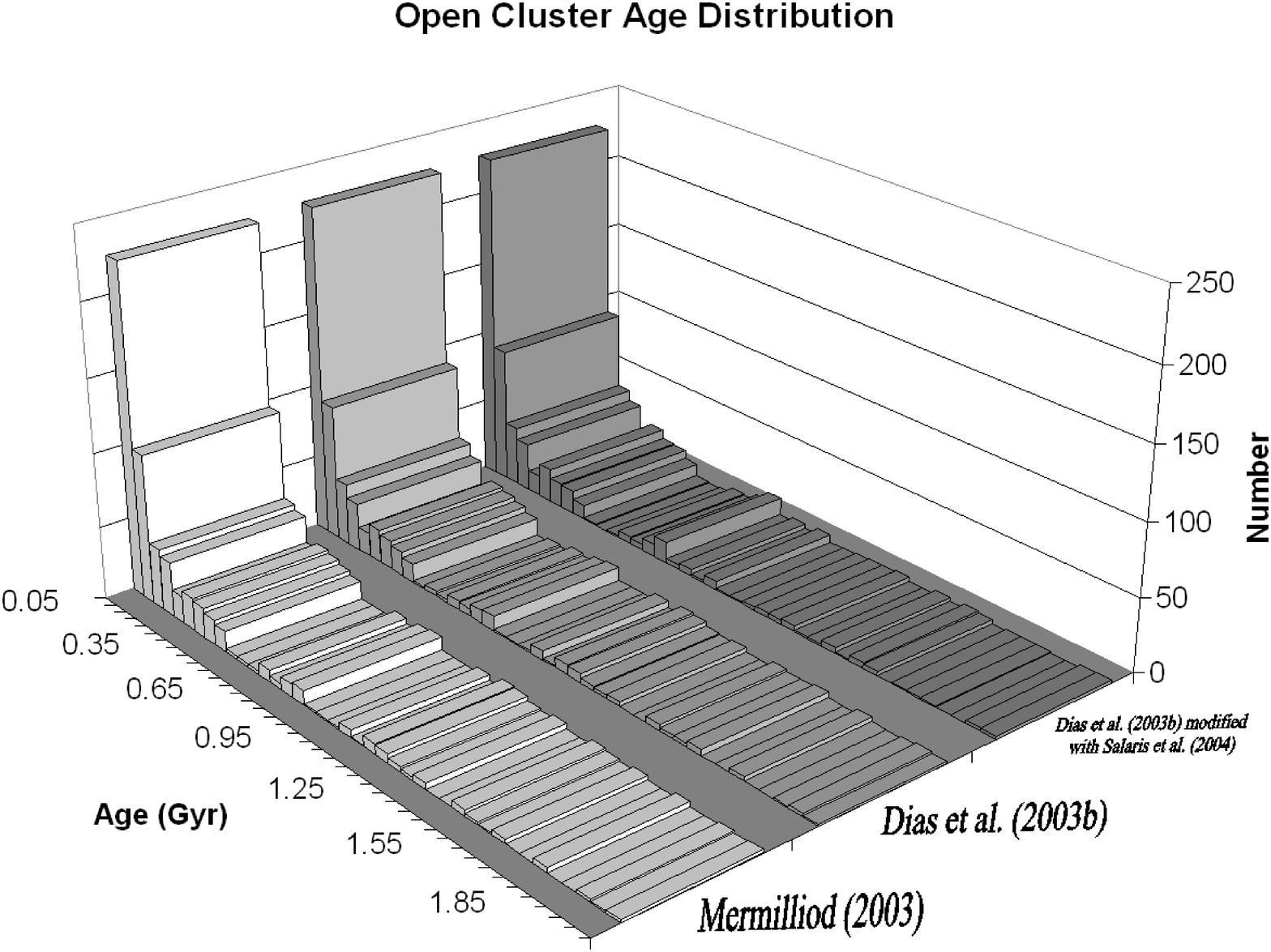,height=16cm,width=14cm,angle=0}
        }}
   \caption[]{Open cluster age distribution from the last update
              (October 2003) of WEBDA (Mermilliod, 2003), white,
              the latest version (October 2003) of the {\it New Catalogue of 
              Optically Visible Open Clusters and Candidates} (Dias 
              et al., 2002, 2003a), light grey, and the Catalogue modified
              with new data from Salaris et al. (2004), dark grey.
             }
      \label{ay}
\end{figure*}

    In this paper we use the three subsamples presented above to construct 
    an age distribution diagram. As explained below, we choose 2 Gyr in 
    order to obtain statistically relevant results. Our samples contain 
    objects from dense, very young clusters to old and almost dissolved 
    moving groups. The age distribution diagram for the three open cluster
    samples considered is shown in Fig. \ref{ay}. Although a detailed 
    analysis of the different features present in this histogram is 
    postponed to Section 4, the figure suggests that several epochs of
    enhanced star formation may have taken place during the last 2 Gyr.

    The age-distance diagrams for the three samples appear in Figs. 
    \ref{mermilliod}, \ref{dias}, and \ref{salaris}. These figures indicate 
    that the open cluster data used in this paper provide an incomplete 
    sample; unfortunately, it is difficult to estimate the degree of 
    incompleteness of this sample due to the irregular distribution of 
    Galactic open clusters. The sample appears to be rather incomplete for 
    open clusters more distant than $\sim$ 3.5 kpc, therefore our conclusions 
    can only be rigorously applied to the solar circle: an annulus of 
    Galactocentric distance $R_{\odot} \ \pm 1.75$ kpc, where $R_{\odot}$ is 
    the solar Galactocentric distance ($\sim$ 8.5 kpc, e.g., Zombeck, 1990).  
    On the other hand, the Hipparcos Catalogue contains numerous selection 
    biases which are difficult to characterize. It is, however, reasonable to 
    assume that the sample is fairly complete (166 clusters) for objects in 
    the solar neighbourhood (distances to the Sun of 1 kpc or less) although 
    selection effects are likely to be very important for clusters located in 
    the outskirts of the Galactic disk as well as in the region behind the 
    Galactic center and for small clusters with Galactocentric distance 
    $ < 8.5$ kpc as they are projected against the dense stellar fields of 
    the southern Milky Way. On the other hand, intermediate age and old open 
    cluster samples are incomplete as clusters with higher $|z|$ are easier 
    to detect than clusters with lower $|z|$. In this paper, we are 
    interested in studying only the recent star formation history of the 
    Milky Way ($\leq$ 2 Gyr), therefore any effects arising from lack of 
    completeness of the sample are likely to be less important than those 
    from age errors. Besides, open cluster samples are biased toward objects 
    where the total luminosity is dominated by ultra-luminous O and B stars 
    (young clusters) or red (or blue) giants (old clusters) as these systems 
    are easier to identify. However, if O and B stars were the only stars the 
    cluster would not be stable since the massive stars are destined to become 
    supernovae, returning their mass to the interstellar medium. Hence, the 
    existence of low mass stars is critical for the survival of open 
    clusters. Old, dynamically depleted open clusters are very difficult to 
    identify although they may exist in large numbers (de la Fuente Marcos, 
    1998). Faint, population depleted open clusters are very difficult to 
    identify against the field stars. Therefore we are working with an 
    incomplete, volume-limited sample that is also flux-limited. 

%
\begin{figure*}
        \centerline{\hbox{
        \psfig{figure=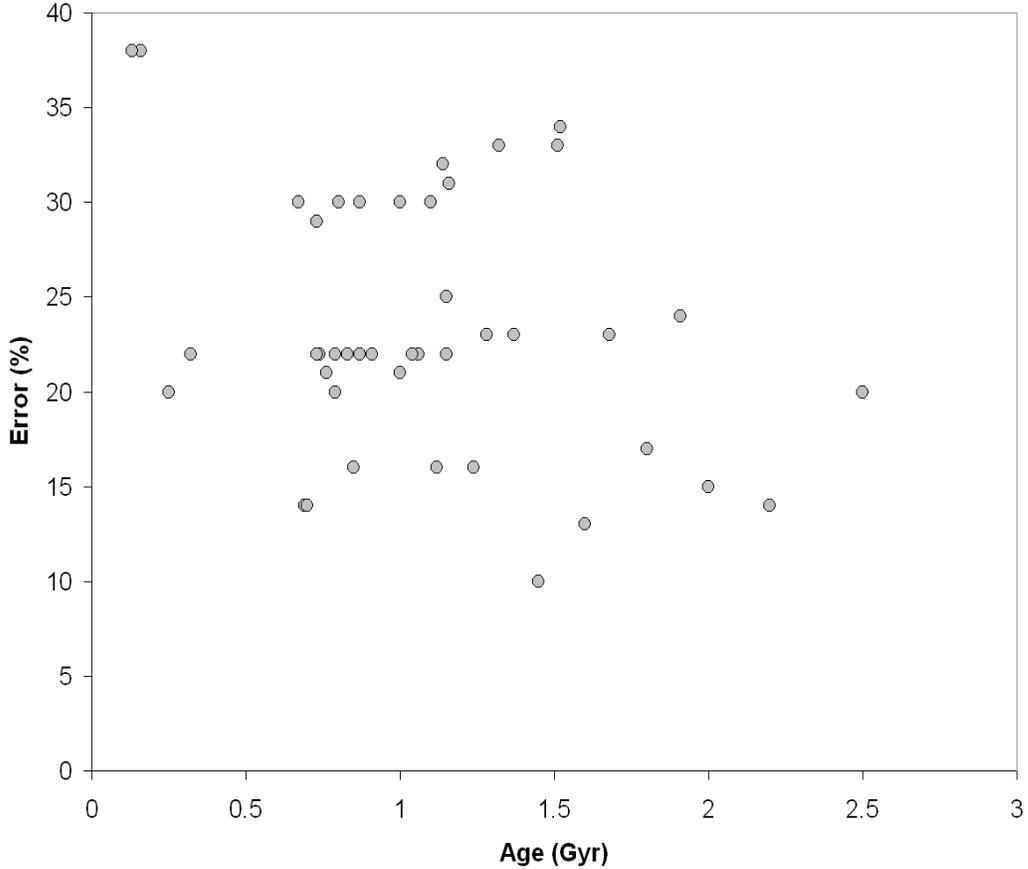,height=12cm,width=14cm,angle=0}
        }}
   \caption[]{Age errors as a function of the open cluster age
              for a selected sample of 52 objects with both age and
              error published (see the text). The average 
              uncertainty is about 22\% and the plot suggests
              that larger errors are associated with younger ages.
             }
      \label{ocerror}
\end{figure*}

%
%
    On the other hand, and as for any other physical parameter, open 
    cluster age determinations are affected by errors. In a large sample, 
    these errors are very likely to be non-homogeneous as different methods 
    have been used by different authors to calculate the ages. The disparity 
    in open cluster ages (when considering different authors, see WEBDA for 
    multiple examples) partly reflects the critical dependence of cluster 
    isochrone fitting on the adopted reddening (often large for open 
    clusters), even small reddening errors can create significant errors in 
    the derived age and metallicity. However, determining effective 
    temperatures and metallicities for cluster turn off stars directly 
    through echelle spectroscopy is free from systematic errors and subject 
    only to uncertainties in the model atmospheres. Some ages for clusters 
    in the samples considered have been determined using the first technique, 
    but some others have been found using the second one or even other 
    indirect methods. It is relatively difficult, specially in papers older 
    than about 5 years, to find published estimations of the errors 
    associated with open cluster age determinations. In order to evaluate the 
    impact of age errors in our analysis we have compiled errors for 52 
    objects within the studied age range from published sources: NGC 6134 
    and NGC 3680 (Bruntt et al., 1999); NGC 2141 (Carraro et al., 2001);
    NGC 2158 (Grocholski \& Sarajedini, 2002); NGC 2112 (Carraro et al., 
    2002); Berkeley 104, Berkeley 60, King 15, NGC 381, Berkeley 64, King 6,
    NGC 1348, Berkeley 23, NGC 2259, NGC 2304 (Ann et al., 2002); NGC 1663
    (Baume et al., 2003); NGC 3990 (Prisinzano et al., 2004); Pismis 19
    (Carraro \& Munari, 2004); IC 166, NGC 752, King 5, NGC 1254, NGC 1278,
    NGC 1817, NGC 2158, NGC 2194, NGC 2192, NGC 2236, NGC 2266, Berkeley 30,
    NGC 2324, NGC 2354, NGC 2355, NGC 2360, Haffner 6, Melotte 71, Pismis 2,
    NGC 2660, NGC 2849, NGC 3680, NGC 4815, NGC 5822, IC 4651, IC 4756,
    Berkeley 42, NGC 6802, NGC 6827, NGC 7044, NGC 2477, NGC 7789, NGC 2204,
    Hyades, and Praesepe (Salaris et al., 2004). Fig. \ref{ocerror} shows the 
    uncertainty as a function of the age for this open cluster sample. The
    average age error is about 22\% and the plot suggests that younger ages
    are affected by larger errors (for a recent discussion on the problems
    associated with age determinations in young open clusters see, e.g., 
    Piskunov et al., 2004). If this error sample can be considered
    as representative of the error range for the entire sample, the 
    youngest cluster (age $<$ 0.5 Gyr) errors are very likely in the range
    50-150 Myr with for older cluster errors in the range 150-250 Myr.
%
%
%
%

       In order to test the reliability of the features found in the
       open cluster age distribution we have generated synthetic age 
       distribution diagrams from the original data. Each simulation 
       was composed of the following steps:

       1) The original sample from Mermilliod (2003) is the input

       2) A randomly distributed error within a given percentage of
          the input value is added/subtracted to each value in the original
          sample.

       3) The simulated open cluster sample is binned at 0.050 Gyr
          intervals. In this way the real age of the open clusters
          in the sample is shifted randomly according to a given
          average error (15, 25, and 35 \%). 

\begin{figure*}
        \centerline{\hbox{
        \psfig{figure=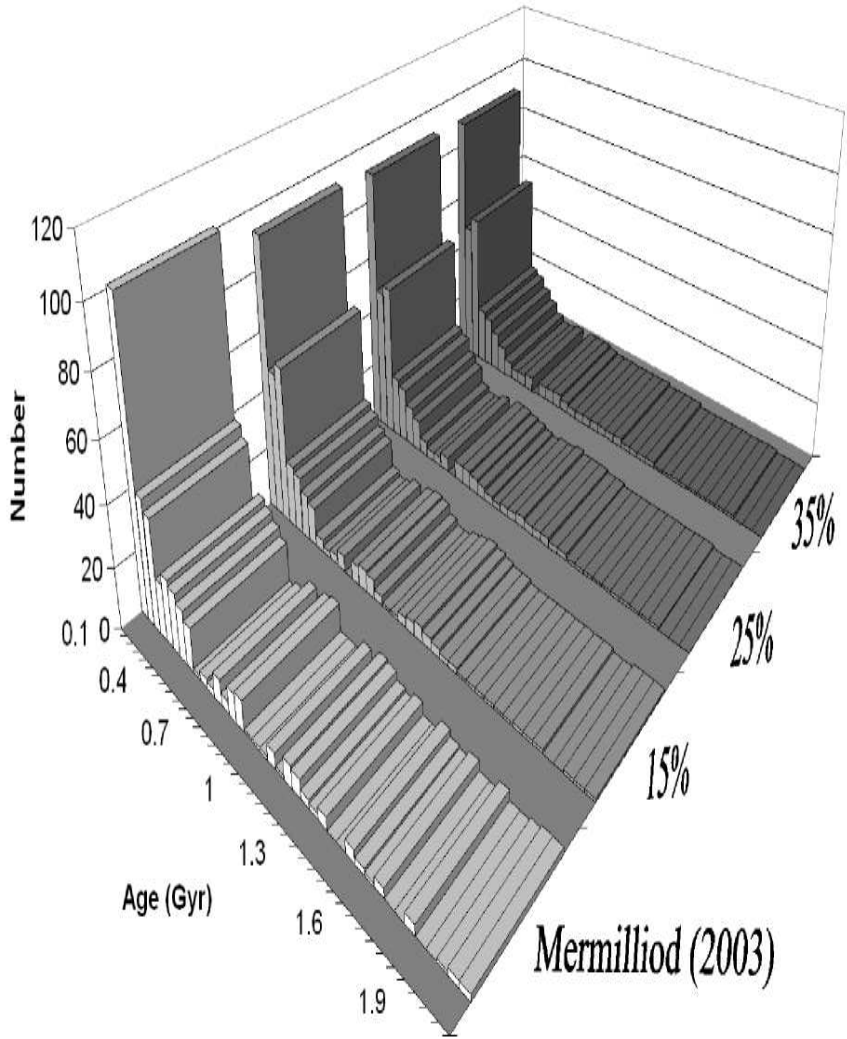,height=14cm,width=14cm,angle=0}
        }}
   \caption[]{Age distribution diagram of 
              the open cluster sample from WEBDA. Each inner
              bar diagram gives the average of 175,000 simulated open 
              cluster samples with different error range (15\%, 25\%,
              and 35\%). 
             }
      \label{error}
\end{figure*}

       4) The process is repeated 175,000 times per error value and
          we calculate the average number of objects per age bin.
          The final, averaged, sample is plotted in Fig. \ref{error}.

       We have carried out three sets with 175,000 simulations to study
       this effect. These simulations appear to suggest that for age
       errors of about 15\% or lower, most of the features in the age 
       distribution remain relatively unaltered. For errors in age 
       determination larger than about 25\% only the youngest features 
       are recovered. 
%
%
%

 \section{The life span of open cluster models} \label{new}
    Fig. \ref{ay} is nothing more than the actual age distribution, a
    raw histogram of open cluster numbers vs. age, of relatively young, 
    nearby open clusters. In order to interpret this distribution
    additional tools are needed. Experimental Stellar Dynamics may help 
    us in this analysis by matching cluster dynamics and cluster data to 
    provide a better view of the star formation history of the Milky Way disk.
    This section summarizes some relevant results on the evolution and
    dissolution of realistic $N$-body star cluster models. Some technical
    details on the modeling process are also discussed as well as 
    the initial mass functions used in the calculations. After an open 
    cluster is formed and its massive stars have expelled the gas it will 
    evolve steadily, losing stars on a relatively long time-scale. Distant 
    encounters among cluster stars together with the Galactic tidal field 
    cause the slow and gradual evaporation of the cluster. Close encounters 
    and stellar evolution can speed up the process, specially for small, 
    sparsely populated clusters. Primordial binaries are also very important 
    (although less than stellar evolution) as they dominate completely the 
    dynamical evolution of poor clusters. Violent, very fast cluster 
    disruption can be induced during encounters with giant molecular clouds 
    (Terlevich, 1983, 1985, 1987). The disintegration time-scale of open 
    clusters depends on both population and Galactocentric distance; but, for
    a given Galactocentric distance, the initial population ($N_o$) as well
    as the initial mass function are the main physical parameters to consider 
    in order to analyze the long term evolution of the cluster.

    The numerical results discussed here come from an extensive set of
    simulations (de la Fuente Marcos, 1995, 1996a, 1996b, 1997a, 1997b; 
    de la Fuente Marcos \& de la Fuente Marcos, 2000, 2002) calculated
    with the standard $N$-body code {\small NBODY5} (Aarseth, 1985, 1994,
    1999, 2003) for clusters located in the solar neighbourhood. These 
    calculations include the effects of stellar evolution, the Galactic 
    tidal field, primordial binaries, and realistic initial mass functions 
    (hereafter IMF). {\small NBODY5} consists of a fourth-order 
    predictor-corrector integration scheme with individual time steps. 
    It utilizes an Ahmad-Cohen (1973) neighbour scheme to facilitate 
    calculation of the gravitational forces, and handles close encounters 
    via two-, three-, four-, and chain regularization techniques 
    (Kustaanheimo \& Stiefel, 1965; Aarseth \& Zare, 1974; Mikkola, 1985; 
    Mikkola \& Aarseth, 1993). Five different IMFs were used to generate
    stellar masses. Two standard power law models (Salpeter, 1995; Taff,
    1974) and three modern IMFs due to the Miller \& Scalo (Miller \& Scalo, 
    1979; Eggleton et al., 1989), Kroupa (Kroupa \& Tout, 1992; Kroupa et 
    al., 1990, 1991, 1993) and Scalo (Scalo, 1986). The IMFs used in our 
    calculations differ mainly in the number of low-mass and high-mass 
    stars generated. The Salpeter IMF is given reasonably well by the 
    approximation
    \begin{equation}
       \xi(M) \approx 0.03 \ M^{-2.35} \,,  
    \end{equation}
    where $\xi(M) \ {\rm d}M$ is the number of stars in the mass interval 
    $M$ to $M \ + \ {\rm d}M$. The Taff IMF is given by 
    \begin{equation}
       \xi(M) \approx 0.03 \ M^{-\alpha} \,,  
    \end{equation}
    with $\alpha = 2.50$ for clusters with $N \leq 100$ and $\alpha = 2.65$ 
    for clusters with $N > 100$. For generating masses with the Miller 
    \& Scalo IMF we use the function (Eggleton et al., 1989)
    \begin{equation}
       M({\rm X}) = \frac{0.19 \ {\rm X}}{(1 - {\rm X})^{0.75} + 
		          0.032 \ (1 - {\rm X})^{0.25}} \,,
    \end{equation}
    where X is uniformly distributed in the interval $[0,1]$. For X $\sim$ 
    0.7--0.999, this gives a power-law expression with $\alpha = 7/3$ and 
    for X $<<$ 1 it gives an approximately constant value. For X $>$ 0.999, 
    the IMF slope is increased to 5. The Kroupa IMF is given by (Kroupa et 
    al., 1993)
    \begin{equation}
       \xi(M) = \left\{ 
	                \begin{array}{ll}
	                   0.035 \ M^{-1.3} \ & 
			     \mbox{if 0.08 $\leq$ M $<$ 0.5} \,, \\    
	                   0.019 \ M^{-2.2} \ & 
			     \mbox{if 0.5 $\leq$ M $<$ 1.0} \,,  \\
                           0.019 \ M^{-2.7} \ & 
			     \mbox{if 1.0 $\leq$ M $<$ $\infty$} \,.
	                \end{array}
		\right.
    \end{equation}
    We use the mass-generating function 
    \begin{equation}
       M({\rm X}) = 0.08 \ + \ \frac{\gamma_{1} \ {\rm X}^{\gamma_{2}} 
				 + \gamma_{3} \ {\rm X}^{\gamma_{4}}}
						{(1 - X)^{0.58}} \,,
    \end{equation}
    as a convenient representation of these relations, where X is uniformly 
    distributed in the interval $[0,1]$ and the $\gamma$-parameters are 
    given by $\gamma_{1} = 0.19$, $\gamma_{2} = 1.55$, $\gamma_{3} = 0.050$ 
    and $\gamma_{4} = 0.6$. No masses less than 0.08 $M_{\odot}$ are 
    generated by this formula. It is based on Scalo's (1986) initial mass 
    function. For the Scalo IMF we use the mass-generating function
    \begin{equation}
       M({\rm X}) = \frac{0.3 \ {\rm X}}{(1 - X)^{0.55}} \,,
    \end{equation}
    where X is uniformly distributed in the interval $[0,1]$. 
    It is similar to Kroupa IMF except for $M <$ 0.16; Scalo IMF gives a 
    smaller number of stars with those masses. The Kroupa IMF is the most 
    recent IMF we have used, it generates a greater number of low-mass stars 
    and a smaller number of heavy stars. On the contrary, the Salpeter IMF 
    gives a smaller number of low-mass stars and a greater number of massive 
    stars. Among these, the Taff IMF gives a smaller number of 
    massive stars than the Salpeter one but greater than the Miller \& Scalo 
    one. The Scalo IMF is intermediate between Miller \& Scalo and Kroupa 
    IMFs. For low-mass stars, the Miller \& Scalo IMF gives a greater number 
    than the Salpeter and Taff ones but smaller than Scalo and Kroupa IMFs. In 
    addition, three different star densities (uniform, $\propto$ 1/$r^2$, 
    Plummer) have been used to generate the initial positions of the stars in 
    the cluster model. Since the initial membership of a star cluster is the
    main parameter with regard to cluster life-time for a given 
    Galactocentric distance, stellar evolution (through the IMF) is the 
    only process able to induce a differential behaviour as a function
    of $N$. 

\begin{figure*}
        \centerline{\hbox{
          \psfig{figure=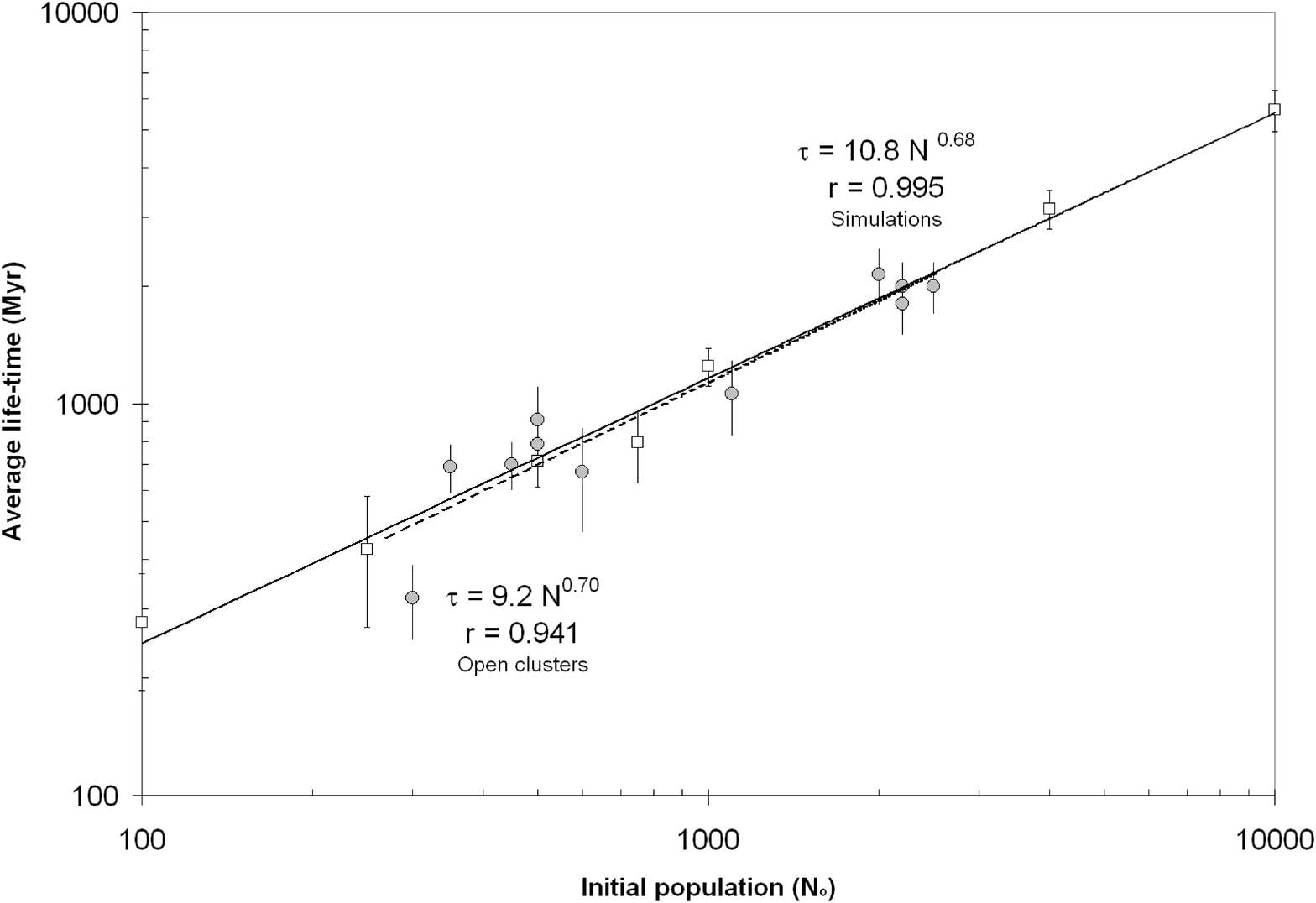,height=14cm,width=14cm,angle=0}
        }}
   \caption[]{Characteristic life-time ($\tau$) as a function of the
              initial membership for a set of realistic open cluster
              models (de la Fuente Marcos 1995, 1996a, 1996b, 1997a,
              1997b; de la Fuente Marcos \& de la Fuente Marcos 2000, 
              2002). The error bars indicate the standard deviation 
              found for each value of $N$. The solid line represents 
              the power law fit to the models data (empty squares), 
              $\tau = 10.8 \ N^{0.68}$ with a correlation coefficient 
              of $r = 0.995$. Actual open clusters are plotted as grey
              circles. The dashed line represents the power law fit
              to the open cluster data (grey circles), 
              $\tau = 9.2 \ N^{0.70}$ with a correlation coefficient
              of $r = 0.941$. Open cluster memberships have been estimated
              using the model described in the text.}
      \label{our}
\end{figure*}

    Averaging cluster life-times for the entire library of models as a 
    function of $N$ to fit a power law to the data we obtain 
    $\tau = 0.011 \ N^{0.68}$ (in Gyr) with a correlation coefficient of 
    $r = 0.995$ (Fig, \ref{our}). This slope is very similar to the slopes 
    found by Baumgardt (2001) and Baumgardt \& Makino (2003) although 
    the details of our simulations are rather different. The above 
    evolutionary track is only strictly valid for cluster models with $N >$
    100-150 members. Unfortunately, for smaller clusters the characteristic
    life-time is not very well defined as the energy released in a single
    supernova event may be larger that the binding energy of the entire
    cluster (de la Fuente Marcos, 1993). For clusters with $N < 50$, 
    life-times are in the range 5-40 Myr and for $N$ = 75, $\tau <$ 80. In 
    general, the fluctuation range for the life-time of small-$N$ cluster
    models is very large and depends strongly on the initial mass function,
    the spatial distribution, and the binary fraction. For richer clusters,
    simulations indicate that, in the solar neighbourhood and neglecting
    interactions with giant molecular clouds and the effect of the cluster
    gas on the early stages of the evolution of the cluster, an open cluster
    has to include about 200-400 stars (at least) in order to survive for 
    about 0.5 Gyr, 400-700 to last 0.7 Gyr, 700-1000 to be detectable after 
    0.9 Gyr and 1000-2000 to survive for about 1.3 Gyr. If a cluster is still 
    observable after 2 Gyr, its initial population was at least 3000 stars. 
    In terms of cluster masses and for an average stellar mass of 
    0.4 $M_{\odot}$, the oldest open clusters considered in this paper 
    (2 Gyr) were born with masses in excess of 1.2 $\times \ 10^3 \ M_{\odot}$
    and very likely around $2 \ \times \ 10^3 \ M_{\odot}$.
%
%
%
%

    The astrophysical interpretation of the solid line in Fig. \ref{our}
    is rather clear: it is the open cluster {\it death line}. Open clusters
    well below the line are young and they evolve towards the line as they
    lose stars into the field. As it evolves, an average open cluster 
    spends most of its life-time close to the {\it death line} moving 
    parallel to it, in the population vs. age diagram. Open clusters become
    part of the stellar field when they cross the line. It is, however, very
    difficult to find open cluster memberships and masses among the 
    published literature. Star counts are in many cases unreliable and
    proper motions are not always available to confirm membership. There is 
    however an observational parameter that may help in open cluster
    membership determinations: the integrated magnitude of the cluster.
    
    The integrated absolute visual magnitude of a star cluster can be 
    written as:
    \begin{equation}
       M_{v}({\rm cluster}) - M_{v}({\rm sun}) = -2.5 \ \log 
               \frac{L({\rm cluster})}{L({\rm sun})} \,,
                 \label{ivm}
    \end{equation} 
    where the absolute visual magnitude of the Sun is 4.82 (e.g. Zombeck, 
    1990) and the luminosity of the cluster could be, in principle, 
    estimated using the mass-luminosity relationship 
    $L \propto M^{\beta}$, with $\beta \approx 3.8$ (Popper, 1980) for
    stars in the main sequence. Unfortunately, the main contribution to the 
    intrinsic brightness of a relatively young cluster is not coming from 
    the stars on the main sequence but from red (or blue) subgiants or
    giants. Taking this fact into account the luminosity of the cluster
    can be written as:
    \begin{equation}
       L({\rm cluster}) = N_{NMS} \ <L_{NMS}> + N_{MS} \ <L_{MS}> \,,
                           \label{oclu}
    \end{equation}
    where $N_{MS}$ and $N_{NMS}$ are the number of stars in the main
    sequence and outside the main sequence, respectively. Their relative
    contribution to the integrated luminosity of the cluster is obtained
    multiplying by the corresponding average luminosity. This average
    luminosity can be obtained from the average mass and the 
    mass-luminosity relationship for the contribution from the main 
    sequence. For subgiants or giants it is also possible to obtain an
    average luminosity using data for standard stars (e.g. Zombeck, 1990).
    In our calculations we will consider an average luminosity for stars
    outside the main sequence of 565 $L_{\odot}$. Combining Eqs. \ref{ivm},
    \ref{oclu}, $N = N_{MS} + N_{NMS}$, and considering the average stellar
    mass $M$ we obtain:
    \begin{equation}
       N = \frac{1}{M^{3.8} + \frac{N_{NMS}}{N} \ <L_{NMS}(L_{\odot})>}
                  {\rm\large 10}^{- 
                    \frac{(M_{v}({\rm cluster}) - M_{v}({\rm sun}))}{2.5}}\,.
                     \label{Npop}
    \end{equation}
    The fraction of stars outside the main sequence can be estimated
    assuming an IMF and taking into account that an estimation of the
    life-time in the main sequence for a given star of mass $M$ and 
    luminosity $L$ is $\propto M/L$. The relationship given by Eq. \ref{Npop}
    can be calibrated using a few clusters with well established memberships
    (e.g. Hyades, Pleiades). For the age range studied in this work the
    fraction of stars outside the main sequence is in the range 0.5-10\%.
    The final value of $N$ is affected by our choice for this fraction but
    the main contribution to the uncertainty in $N$ is coming from the
    observational determination of the integrated magnitude. For Galactic
    open clusters this uncertainty could be $\pm 0.5$ mag (Battinelli et al.,
    1994; Lata et al., 2002). This translates into 20\% errors in $N$. 
    We have applied this approach to estimate the current population
    of a small sample of open clusters (Hyades, Pleiades, IC 4756, NGC 2204,
    NGC 2506, NGC 7789, NGC 1245, NGC 381, NGC 2324, and NGC 2192) with age, 
    error in age, and integrated absolute visual magnitude available from
    published sources (WEBDA; Battinelli et al., 1994; Gray, 1965; Pandey et 
    al., 1989; Lata et al., 2002; Sagar et al., 1983; Spassova \& Beav, 
    1985). The results are also plotted in Fig. \ref{our}, the dashed 
    line represents the power law fit to the open cluster data (grey 
    circles), $\tau = 9.2 \ N^{0.70}$ with a correlation coefficient of 
    $r = 0.941$. Only one open cluster younger than 500 Myr has been
    included because all of them are well below the limiting line. The power
    law fit includes all the plotted points. If the youngest cluster
    is not considered then the correlation coefficient is essentially 1.
    In any case, Fig. \ref{our} appears to confirm that observational data
    and results from $N$-body simulations are fully consistent within
    the error limits.

 \section{Method and results}
    In this section we provide a detailed description of our open cluster
    age distribution method. As stated in Section 2, our results are mainly
    sensitive to the errors in age determination but also to the degree of 
    completeness of the open cluster sample. These two effects have been 
    analyzed in Section 2 and are discussed in detail below. On the other
    hand, any errors in theoretical stellar models also propagate into the
    results obtained because the published open cluster ages always make
    reference to theoretical stellar models, sometimes through direct 
    isochrone comparison, others through the use of morphological features
    found in the cluster color-magnitude diagram.
 
    \subsection{Method and assumptions}
       The open cluster catalogues offer high-quality data for a 
       relatively large number of objects in the Milky Way disk, both in 
       the solar neighbourhood and beyond. In particular, the age data 
       can be used to construct an age distribution for the Milky Way open 
       clusters, Fig. \ref{ay}. Assuming that the sample considered is
       representative of the entire Galactic disk, the star formation rate
       can be derived from its age distribution, as the number of open 
       clusters in each age bin is, by hypothesis, correlated with the number 
       of open clusters initially born at that time. Once an age distribution 
       is available, it is in principle easy to recover the star formation 
       history that gave rise to the observed age distribution by using 
       results from realistic $N$-body simulations. This method permits the 
       reconstruction of the global star formation history of the 
       Galactic disk with a time resolution of 0.050 Gyr over the last 
       2 Gyr. Our method does not assume any a priori structure or 
       condition on the star formation rate and it 
       basically consists of two steps: (i) Construct a representative 
       sample of open clusters. In principle, the optimal approach will 
       be to construct a volume-limited open cluster sample in the solar 
       vicinity (see Section 6). The cluster sample spans a very large 
       range of ages. (ii) Construct the age distribution diagram for the 
       sample. (iii) Infer the star formation history from the diagram.
       Observations of bursts of star formation in other galaxies indicate 
       that intense star formation is always associated with production of
       large star clusters. The life span of larger clusters is longer; 
       therefore, an usually high number of open clusters at a given time 
       interval can be interpreted as the result of an event of enhanced 
       star formation at that given age. This is probably the main a priori
       assumption maintained throughout the paper although it appears to be
       strongly supported by observational evidence. If star cluster masses 
       are sampled from an open cluster initial mass function, larger numbers 
       translate into increased probability of formation of large star 
       clusters. Luminous, and therefore rich, young star clusters are found 
       whenever there is vigorous star formation, whether it be in galaxy 
       mergers or starburst galaxies. Galaxies with very active star formation
       have proportionally more of their stars in clusters than in the field,
       with some of them devoting as much as 15-20\% of their luminosity
       to clusters (Larsen \& Richtler, 2000). Presently available 
       observational data strongly suggests that the cluster forming
       frequency is highest during violent bursts of star formation 
       (van den Bergh, 2000). 
        
    \subsection{Results} \label{history}
\begin{figure*}
        \centerline{\hbox{
        \psfig{figure=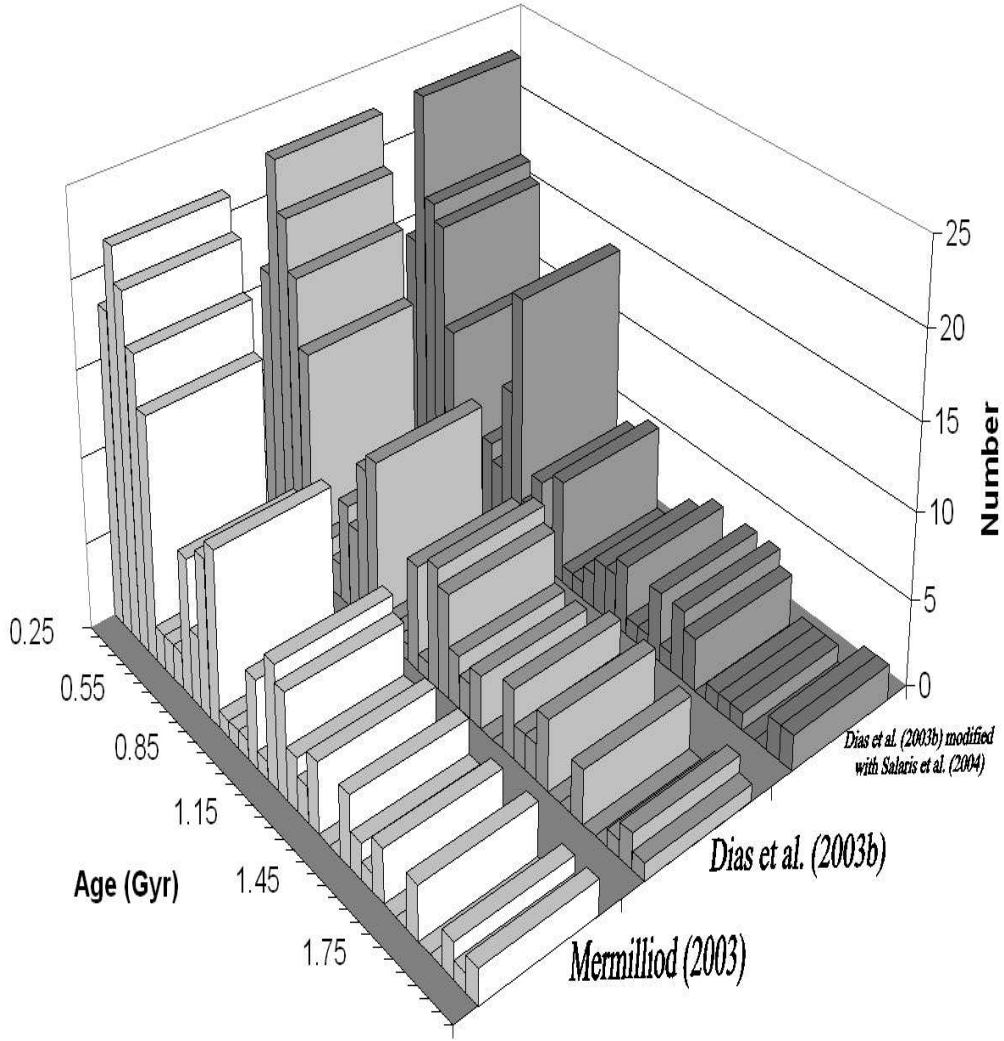,height=14cm,width=14cm,angle=0}
        }}
   \caption[]{Detail of Fig. \ref{ay} for the age range 0.2-2.0 Gyr,
              same bin size.
             }
      \label{ayz}
\end{figure*}

       The resulting star formation history comes directly from the age
       distribution (Figs. \ref{ay}, \ref{ayz}), in an approach which 
       assumes that the most frequent ages of the open clusters indicate 
       the epochs when the star formation was more intense if we take into
       account that star clusters are the elementary units of the star 
       formation process. We have not included any volume correction 
       because the number of clusters more distant than about 6 kpc is 
       negligible for the two samples considered. The volume effect as well
       as evolutionary corrections are considered in Section 6). We refer 
       basically to the clumps of clusters at about 0.35, 0.70, 1.13, 1.50, 
       and 1.93 Gyr. These clumps will be identified as burst 1, 2, 3, 4, 
       and 5, respectively. The actual existence of burst 1, 2 and 3 is not 
       significantly (see the detailed discussion below) affected by 
       uncertainties in the cluster ages as errors are smaller than 0.3 Gyr 
       for all clusters (younger than 1.2 Gyr) in the samples considered. 
       However, for clusters older than 1.2 Gyr age errors are larger, 
       therefore the statistical significance of bursts 4 and 5 is not, in
       principle, as well established as for 1, 2, and 3.    

%
%
%
%
  
       On the other hand, if we assume that the Galactic disk is not 
       currently undergoing a burst of star formation and that the age 
       distribution for stars younger than 0.2 Gyr is representative of 
       the monotonic (quiescent) star formation rate in the Milky Way 
       disk (i.e. it has not been polluted by an enhanced episode of 
       star formation), then the fraction of open clusters that survives 
       for up to about 0.25 Gyr is about 92\%. This corresponds to the 
       characteristic life-time of clusters with $N \leq 150$. In other 
       words, observations indicate that only 8\% of all the open clusters 
       are born with stellar populations larger than about 200 stars or 
       masses $> 80 \ M_{\odot}$. This feature can be interpreted as 
       implying that star formation takes place preferentially in clusters 
       of that size. This result, that most of the stars appear to have been 
       formed in small star clusters, has already been pointed out by Kroupa 
       (1995a, b, c) and de la Fuente Marcos (1997b). However, a single 
       3,000 stars open cluster can produce as many field stars as 60 small 
       (50 stars) clusters and, as pointed out above, the number of large 
       clusters violently destroyed during cluster-giant molecular cloud 
       interactions, even though difficult to estimate, can be very 
       significant. On the other hand, during an episode of enhanced star
       formation even much larger clusters (up to 40,000 stars) can be formed.
       Therefore it may also be possible that most of the field stars have 
       been formed in large clusters, however we do not observe many large 
       clusters today and the previous possibility remains rather speculative. 
       In any case, the cumulative age distribution for the full open 
       cluster sample shows a departure from the predictions of constant 
       formation rate and exponentially declining dissolution rate.

    \subsection{Quiescent vs. burst of star formation}
       Assuming that the Galactic disk has never been polluted by any
       extragalactic stellar populations, our previous analysis suggests
       that superimposed on a relatively small level of quiescent star
       formation, mainly in small-$N$ star clusters, the star formation
       rate has experienced at least 5 episodes of enhanced star formation
       lasting about 0.2 Gyr each, with production of larger clusters. This 
       cyclic behavior appears to show a period of 0.4$\pm$0.1 Gyr. Although
       the observed cyclic behavior can be easily explained as triggered
       by density waves (Hernandez et al., 2000a; Martos et al., 2004), 
       the analysis in this section is also exploring the possibility of 
       periodic, tidally triggered star formation bursts as an alternative
       (or possibly concurrent) scenario to explain the enhancements observed 
       in the open cluster age distribution. There is evidence for a recent 
       pericentric passage of the Sagittarius Dwarf Spheroidal Galaxy $\sim$ 
       50 Myr ago (Johnston et al., 1999). If this close interaction was able 
       to tidally trigger star formation in the disk of the Milky Way, its 
       effects should currently be observable and it could be considered as 
       burst 0. There are in fact observational signatures of unusually rich 
       young open clusters (or even protoglobulars) in the disk of the Milky 
       Way: e.g. Cyg OB2 (Kn\"odlseder, 2000; Kn\"odlseder et al., 2002; 
       Comer\'on et al., 2002; Hanson, 2003). We will however neglect this 
       possibility in our subsequent discussion.

       \subsubsection{The 0.35 Gyr burst}
          The most recent star formation bursts are also the most likely
          bursts to have occurred, since they took place in the very
          recent past, and so are less affected by the age errors. 
          Unfortunately, our analysis in Section 2 appears to suggest that
          age errors are larger for younger clusters. The youngest 
          (0.25-0.45 Gyr) burst found in this work appears around the same 
          age in all the three samples, lasting 0.2 Gyr that is the typical 
          duration of bursts found in previous work. This burst is 
          temporally coincident with a perigalactic passage of the Small 
          Magellanic Cloud (hereafter SMC) with the Milky Way as predicted 
          by Lin et al. (1995). Harris and Zaritsky (2004) have studied the 
          star formation history of the SMC and have found four major 
          bursts of star formation: 0.06, 0.4, 2.5, and 8.4 Gyr. Models by 
          Zaritsky and Harris (2004) indicate that the 0.4 Gyr peak in the 
          SMC stellar age function coincides with one of the epochs of 
          closest approach between SMC and the Milky Way. During the 
          interaction, the elevation in the star formation rate was 
          significant ($>$ factor of 2) over the quiescent rate of star 
          formation in the SMC.

       \subsubsection{The 0.70 Gyr burst}
          This burst appears at 0.6-0.8 Gyr, lasting also 0.2 Gyr, in 
          WEBDA and NCOVOCC but appears to be wider, 0.6-1.0 Gyr, for
          the sample including Salaris et al. (2004), shifting the peak
          to 0.8 Gyr. Johnston et al. (1999) have found that the Sagittarius 
          Dwarf Galaxy had a pericentric passage $\sim$ 0.7 Gyr ago. This
          close interaction is temporally coincident with burst 2.

       \subsubsection{The 1.13 Gyr burst}
          From WEBDA, this burst started 1.3 Gyr ago, lasting about 350
          Myr. However, NCOVOCC gives the same starting age but lasting
          0.3 Gyr. The sample corrected with Salaris et al. (2004) data
          differs also only in the duration, 0.2 Gyr. This burst is 
          temporally coincident with a perigalactic passage of the SMC 
          with the Milky Way (Lin et al., 1995).

       \subsubsection{The 1.50 Gyr burst}
          The three samples indicate that the burst started 1.6 Gyr ago,
          lasting 0.2 Gyr. This burst appears to be coincident with a close 
          encounter between the Magellanic Clouds and the Galaxy (Murai 
          \& Fujimoto, 1980; Gardiner et al., 1994; Lin et al., 1995). 
          On the other hand, the Sagittarius Dwarf Galaxy had another 
          pericentric passage around the same time (Johnston, 1998; Johnston 
          et al., 1999).

       \subsubsection{The 1.93 Gyr burst}
          This is the oldest star formation burst identified in this work
          and it is also the most affected by the age errors. WEBDA and 
          NCOVOCC indicate that this burst started 2 Gyr ago, lasting 0.15 
          Gyr. The third sample suggests a burst lasting 0.1 Gyr.
          This is the only burst that does not appear to coincide with any
          predicted or modelled interaction with a neighbour galaxy.

       In spite of the time coincidences, interpreting the observed
       periodicity in the age distribution of young, nearby open clusters 
       as evidence in support of a tidally triggered star formation history 
       is somewhat speculative. On the other hand, a close interaction with 
       the Magellanic Clouds cannot be compared dynamically to that of the 
       Sagittarius or the Canis Majoris dwarf galaxies. 

       Using our method without any corrections it is, however, difficult to 
       estimate the evolution of the absolute star formation rate with the 
       age of the disk. Nevertheless, it is possible to attempt a rough 
       estimate: the number of clusters in bursts 1, 2, 3, 4, and 5 from 
       the WEBDA sample is 73, 31, 28, 10, and 5 respectively, for the 
       Dias et al. sample is 73, 30, 28, 12, and 4 and for Salaris et 
       al. is 66, 49, 12, 12, and 4. Data seem to suggest that for WEBDA 
       and Dias et al. burst 2 was likely the strongest. If under quiescent 
       star formation only 2\% of open clusters survive for more than 2 Gyr, 
       during burst 2 the star formation increased by at least a factor 4. 
       Salaris et al. data also indicate that burst 2 was the most important, 
       followed by burst 4. During burst 2 the number of open clusters 
       formed increased by a factor 8 at least.  
     
    \subsection{Limitations} 
       As pointed out before, the star formation history inferred for the 
       Milky Way disk in the previous section is affected by selection 
       effects and cluster age errors, therefore some features in the 
       derived age distribution could be caused by the incompleteness of 
       the sample. As a result of these limitations, the actual amplitude 
       of the bursts is likely higher. Working with a flux-limited sample 
       makes it very difficult to estimate the evolution of the absolute 
       star formation rate with the age of the disk. Nevertheless, in this 
       section we try to assess the accuracy of our results. 

       The age errors affect considerably the duration of the star formation 
       events, since they tend to scatter the ages of the star clusters 
       originally born in a burst. We can expect that this error could smear 
       out peaks and fill in gaps in the age distribution. In Section 2, we 
       have tested our results using synthetic age distribution diagrams from 
       the original data (see Fig. \ref{error}). Simulations appear to 
       indicate that for age errors of about 15\% or lower bursts 2, 3, 4, 
       and 5 remain relatively unaltered both in terms of age and duration, 
       however burst 1 seems to split into two bursts, one at 0.2 Gyr and 
       another one at 0.5 Gyr. For errors in age determination larger than 
       about 25\% only bursts at 0.2, 0.5, and 0.7 are recovered. However, 
       our error analysis in Section 2 appears to indicate that the older the 
       cluster, the lower the error. Therefore it could be possible that 
       errors for clusters older than 1.5 Gyr are in the range 15-25\% and 
       the five bursts can be recovered.

       Our derivation of the star formation history of the Milky Way disk 
       uses a sample of dynamically bound, disk star clusters or open 
       clusters. This sample is assumed as not polluted by unbound star 
       clusters or associations. However, associations dissolve on a 
       time-scale of $\sim$ 50 Myr so even if some degree of contamination 
       remains, the contribution to the final conclusions is negligible.
       It is only significant for the first bin in our age distribution
       diagram. On the other hand, it is known that a typical cluster in the 
       outer disk will survive about twice as long as one in the inner 
       disk (Janes et al., 1988). The outer disk group is an homogeneous
       group whose members are dissolving on a time-scale on the order of
       3-4 Gyr. This contribution is however negligible as we restrict
       our analysis to clusters younger than 2 Gyr. In our previous analysis 
       we have assumed that this age distribution is depopulated from 
       extragalactic objects, so no open clusters have been incorporated 
       into the galactic disk from accreted galaxies. An alternative, 
       plausible but rather speculative scenario with excess clusters coming 
       from accreted dwarf galaxies is considered in Section 8.

       Limitations on the applicability of our method to a wider time
       range appear in connection with the small number of known open 
       clusters older than 2 Gyr. With a resolution of 0.05 Gyr, our 
       results are not statistically significant for an age older than 
       2 Gyr as the number of objects per age bin is too low. The need 
       to provide reliable results limits the age range over which we 
       can derive the star formation history to 0-2 Gyr, with the time 
       resolution considered throughout the paper. However, our method can 
       be applied to a wider time range if the time resolution is
       lowered (de la Fuente Marcos \& de la Fuente Marcos, 2004).

 \section{The last 0.2 Gyr}
\begin{figure*}
        \centerline{\hbox{
        \psfig{figure=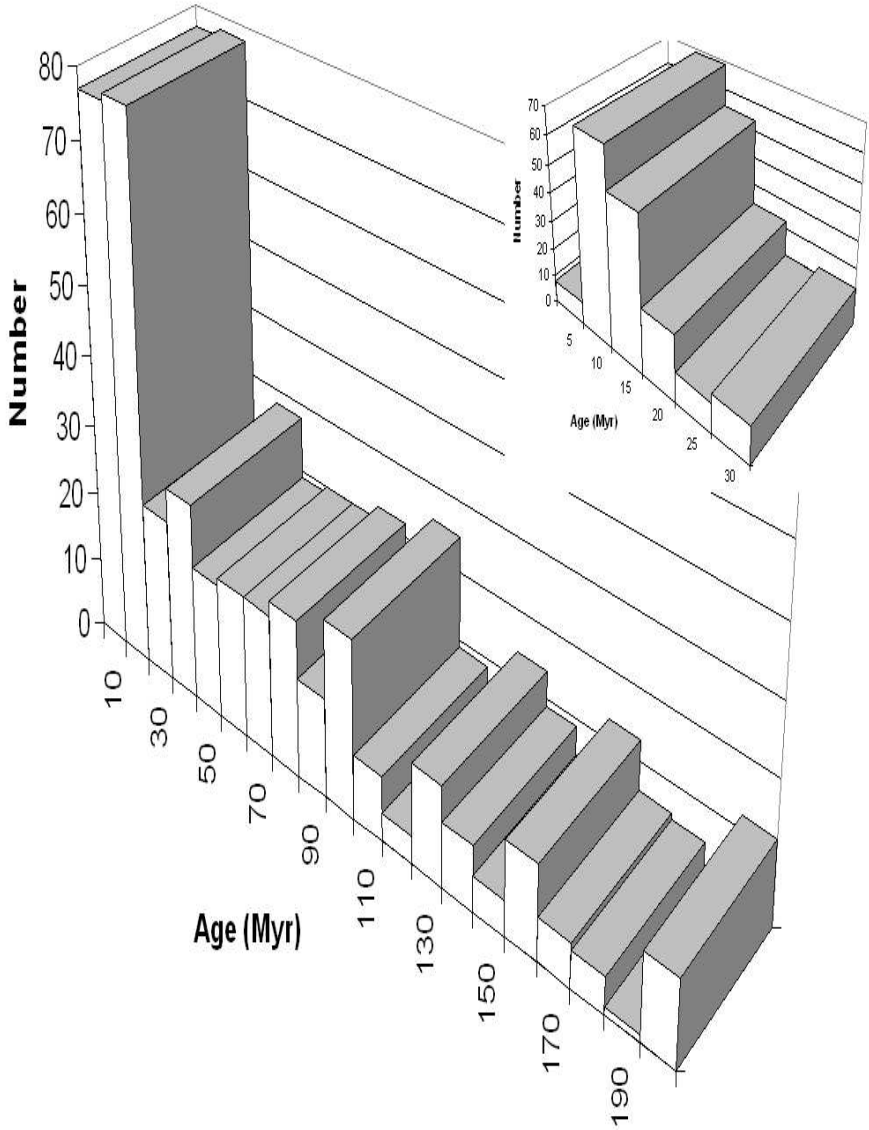,height=16cm,width=14cm,angle=0}
        }}
   \caption[]{Open cluster age distribution from one of the last update 
              (October 2003) of WEBDA, for the last 0.2 Gyr (main panel) 
              with a time resolution of 0.01 Gyr. In the small panel we
              display the last 0.03 Gyr with a resolution of 0.005 Gyr.
            }
      \label{vrec}
\end{figure*}

    Contrary to what is sometimes assumed in the early stages
    of the star formation history of the Milky Way disk, which were ruled 
    by accretion events and early mergers of satellite dwarf galaxies and
    star clusters that contributed to the formation of a substantial 
    fraction of the old and metal-deficient stars in the Galactic disk,
    during the last 0.2 Gyr the star formation rate (see Fig. \ref{vrec})
    has been remarkably stable with apparently no events of enhanced star 
    formation.

%
%
%
%

    Fig. \ref{vrec} appears to indicate that the open cluster sample younger
    than 0.2 Gyr can be considered representative in the analysis of
    quiescent star formation in the Milky Way disk. The characteristic 
    time-scale to assemble and {\it give birth} to an open cluster seems to 
    be $\sim$ 10 Myr with (at least) about 80 new open clusters being formed 
    every 10 Myr or 8 clusters/Myr within about 3 kpc from the Sun. About 
    80\% of newly formed clusters are destroyed during the first 20 Myr of 
    life.
 
    For a time resolution of 5 Myr (Fig. \ref{vrec}, small panel) a very
    interesting feature is visible: clusters younger than 5 Myr are scarce.
    In principle, this feature may be interpreted as a minimum in the star
    formation rate, however young clusters are still embedded in their parent 
    molecular cloud and this makes them difficult to detect. Dust 
    obscuration appears to be the natural explanation for the origin of this 
    feature.  On the other hand, there is a sharp decrease in the number of 
    open clusters observed for ages older than 15-20 Myr. There is an initial 
    very rapid rate of cluster dissolution, gradually declining thereafter. 
    Again the most simple interpretation is to assume that an enhanced star 
    formation period  started $\sim$ 20 Myr ago. However, this feature can be 
    better explained if we consider stellar evolution in open clusters. For 
    a metallicity $Z$ = 0.02, stellar evolution models (e.g. Schaller et al., 
    1992; Bressan et al., 1993) indicate that stars with masses above 
    10 $M_{\odot}$ and older than about 20 Myr have already exploded as 
    supernovae. For a small cluster ($N <$ 75), the energy released during 
    one supernova event can be equivalent to the binding energy of the entire 
    cluster. Therefore, early supernovae appear to be responsible for the 
    violent destruction of about 80\% of young open clusters. 

    In the following section, we use this information to generate a 
    power-law open cluster initial mass function with slope -2.7.

    \subsection{Open Cluster Initial Mass Function} \label{distri}
       The frequency distribution of stellar masses at birth, the so-called 
       stellar initial mass function (hereafter SIMF), is a fundamental 
       parameter to study the stellar mass spectrum. Salpeter (1955) used 
       the observed luminosity function for the solar neighbourhood and 
       theoretical evolutionary times to derive a SIMF which may be 
       approximated by a power-law:
       \begin{equation}
          n(m) \propto m^{-\alpha}\,,
       \end{equation}
       where $n(m)$ is the number of stars per unit mass interval. The 
       original value of $\alpha$ found by Salpeter is 2.35, for masses 
       between 0.4 and 10.0 $M_{\odot}$. The determination of the SIMF is 
       a cornerstone in Astrophysics, since the SIMF determines the 
       evolution, surface brightness, chemical enrichment, and baryonic 
       content of all galaxies and it is also a fundamental link between
       stellar and galactic evolution. For recent reviews on the SIMF see 
       Kroupa (2002) and Chabrier (2003).

       As many of the stars in the disk of the Milky Way appear to be the 
       result of a clustered mode of star formation, the open cluster
       initial mass function (hereafter OCIMF) is also a fundamentally
       important distribution function to study cluster formation, chemical
       evolution of galaxies, and star formation in general. The fact that
       the SIMF and the OCIMF can be intimately related was first pointed 
       out by Reddish (1978). However, deriving the OCIMF is very difficult 
       as cluster mass determinations are affected by several observational 
       and theoretical biases: obscuration by dust, flux limits, incomplete 
       sampling of cluster haloes, lack of information on the evolutionary 
       state of the cluster, and uncertainties in stellar models and cluster 
       dissolution rates. Although the OCIMF is not well constrained it can
       be written as a power law of the type
       \begin{equation}
          N(M) d(M) \sim M^{-\alpha} dM
       \end{equation}
       for $M_{min} < M < M_{max}$, $N(M)$ is the number of open clusters
       per unit mass interval. The cumulative distribution will be 
       \begin{equation}
          f(M) \sim M^{-\alpha+1}\,.
       \end{equation}
       If the process by which open clusters are formed is also able to 
       produce individual stars, it appears reasonable that both SIMF and
       OCIMF should show similar slopes within the observational errors.
       Reddish (1978) found a slope equal to -2.2 from a sample of 72 open
       clusters. V\'azquez and Feinstein (1989) estimated the OCIMF for the 
       Milky Way disk using a sample of 130 clusters from Lyng\aa's (1987) 
       catalogue. In order to find the distribution, they used the values 
       of the masses quoted in the catalogue to obtain a slope 
       $\alpha = 2.74 \pm 0.09$ for all the clusters in their sample. They 
       also found that the slope is different for clusters with Galactocentric
       distances smaller than the Sun's and clusters with larger 
       Galactocentric distances, although this may be an artifact due to
       incompleteness of the data. The topic of the distribution of the 
       masses of star clusters has been re-visited for various star-forming 
       galaxies (e.g. Bik et al., 2003; Elmegreen et al., 2000; Whitmore et 
       al., 1999). Kroupa and Boily (2002) have recently approached the
       subject from a theoretical point of view focusing mainly on large
       clusters. They found that the entire Galactic population II stellar
       spheroid can be generated if star formation proceeded via embedded 
       clusters distributed like a power-law mass function with slope 
       in the range (1.9, 3.6).

\begin{figure*}
        \centerline{\hbox{
        \psfig{figure=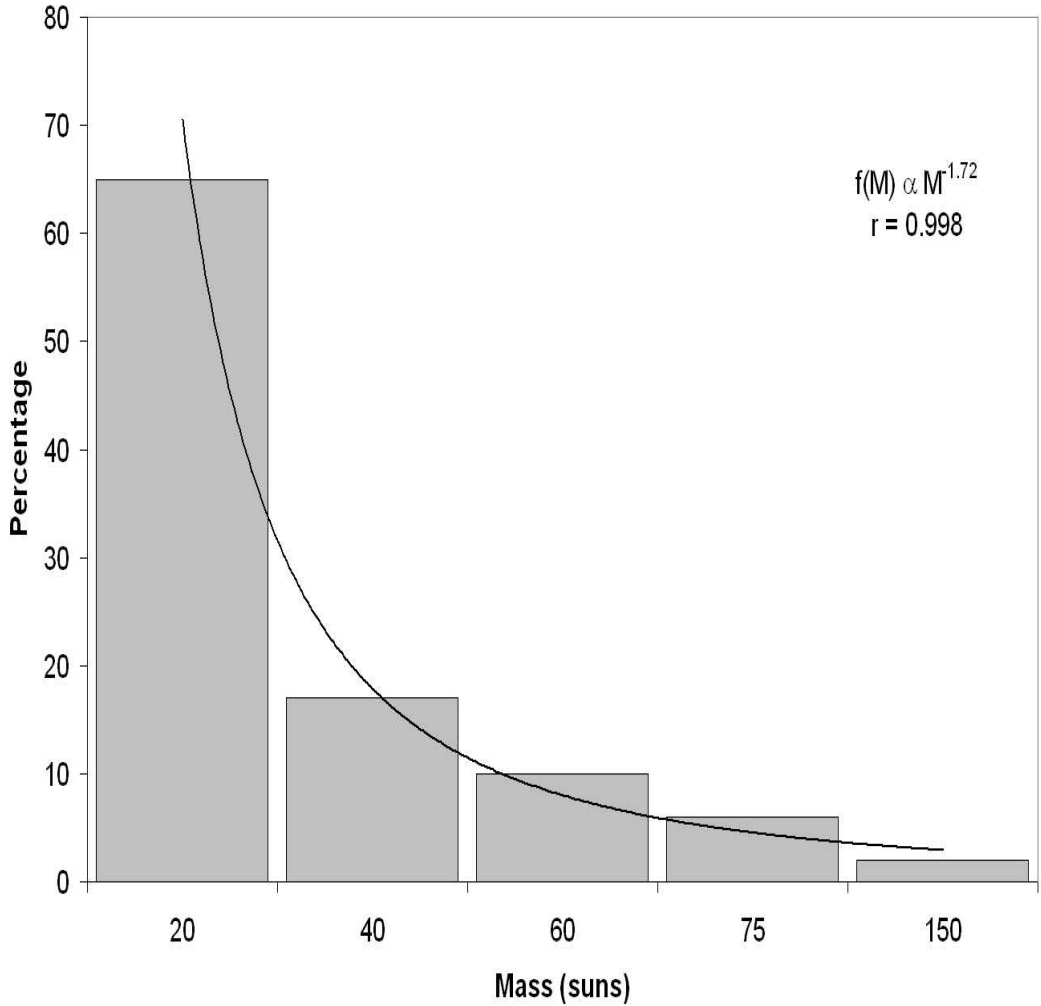,height=14cm,width=14cm,angle=0}
        }}
   \caption[]{Frequency distribution of the masses of open clusters.
              The OCIMF has been constructed from a representative
              sample of open clusters of known age fitting the age to 
              star cluster evolution tracks from $N$-body calculations
              to infer initial cluster masses and creating the 
              distribution from the ensemble of masses obtained. 
             }
      \label{ocimf}
\end{figure*}

       In this section we will infer the OCIMF for the quiescent
       Milky Way disk without considering the values of the cluster masses 
       provided by the catalogues but taking into account the life-time scale 
       obtained from simulations in Section \ref{new} and the open cluster 
       sample described in Section \ref{sample} to derive the OCIMF for the 
       Milky Way disk. This method to construct an OCIMF has been outlined
       in Kroupa and Boily (2002). In Section \ref{history} it has been 
       evidenced that the disk of the Milky Way may have experienced several 
       episodes of enhanced star formation over the last 2 Gyr, this implies 
       that the OCIMF cannot be derived from the total sample of clusters, 
       because the starburst will bias the OCIMF towards massive clusters. 
       However, if we restrict our analysis to clusters with ages less than 
       about 200 Myr, these clusters can be used to derive the OCIMF. This 
       choice implies that the star formation rate has remained almost 
       constant during the last 200 Myr; this assumption appears to be 
       reasonably supported by our previous analysis. For an average stellar 
       mass of 0.4 $M_{\odot}$, we obtain that $\sim$ 65\% of clusters have 
       masses $\le 20 M_{\odot}$, about 17\% with $20 M_{\odot} 
       < M < 40 M_{\odot}$, 10\% with $40 M_{\odot} \le M < 60 M_{\odot}$, 
       6\% with $60 M_{\odot} < M < 75 M_{\odot}$, and $\sim$ 2\% have masses 
       $> 75 M_{\odot}$, with a likely average of 150 $M_{\odot}$. 
       Observations and simulations seem to suggest that 
       $M_{min} \sim 2 M_{\odot}$ ($N_{min} = 5$) and 
       $M_{max} \sim 16,000 M_{\odot}$ ($N_{max} = 40,000$). These 
       results are consistent with an open cluster population being sampled 
       from a power-law OCIMF with $\alpha \approx 2.7$ with a correlation 
       coefficient $r = 0.99$ (see Fig. \ref{ocimf}). The error associated 
       with this determination can be estimated from the standard deviation 
       obtained for the life-time of simulated clusters and is $\sim 20\%$. 
       In this analysis we are not taking into consideration that a number of 
       clusters (those younger than $\sim 60$ Myr) may be still in the 
       evolutionary stage driven by gas expulsion. This simplification is 
       likely to have a minor impact on the overall results from this section 
       because our results indicate that a significant fraction of open 
       clusters formed during quiescence are in fact small and therefore 
       short-lived.
%
%
%
 \section{From the age distribution to the star formation history}
    In the previous sections we have assumed that the open cluster sample
    under study is representative of the entire Galactic disk, therefore 
    the evolution of the star formation rate can be inferred from its age
    distribution, since the number of open clusters in each age bin has
    to be correlated with the number of objects initially at that time
    as a result of dissolution processes. The most reliable transformation
    of the open cluster age distribution into history of the star formation 
    rate comprises two intermediate corrections, namely the volume and
    evolutionary corrections. As stated in Section 4, a scale height
    correction can be neglected because the studied age range is $< 2$ Gyr.
    The applied corrections are explained in the following sections.

    \subsection{Evolutionary correction} 
       As pointed out in Section 3, a correction due to the dynamical 
       evolution of the cluster (cluster disintegration) is needed because
       our sample includes clusters with different initial populations
       ($N_o$). The more massive clusters have a life expectancy higher
       than the short-lived, small-$N$ clusters, thus the latter would
       be missing in the older age bins. It is however possible to correct
       for this effect using the OCIMF obtained in the previous section
       as well as the results from Section 3. 

       The corrections are given by the following formalism. The number
       of clusters born a time $t$ ago ($t$ = 0 is present time) is
       the current number observed at age $t$ divided by the fraction
       of surviving clusters expected for an age $t$, $f$. The fraction 
       of surviving clusters changes with $t$, therefore the life-time
       scale obtained in Section 3 is used to estimate the correct fraction
       of surviving clusters to be applied. In detail: for the second bin
       we have corrected by the same number in bin 1; bins 3-4 have been
       corrected using $f$ = 0.18; $f$ = 0.08 has been used in bins 5-9;
       bins 10-16 have been corrected by $f$ = 0.02. After 0.8 Gyr, we have
       assumed that no cluster formed during quiescent star formation 
       epochs may have survived; therefore, the correction is made in a
       slightly different way: we add the reference level at $t$ = 0 to
       the standard correction with $f$ = 0.02. The age histogram from
       WEBDA and the corrected distribution appear in Fig. \ref{best}.
       The star formation history outlined in Section 4 is somewhat 
       altered:

%
%
\begin{figure*}
        \centerline{\hbox{
        \psfig{figure=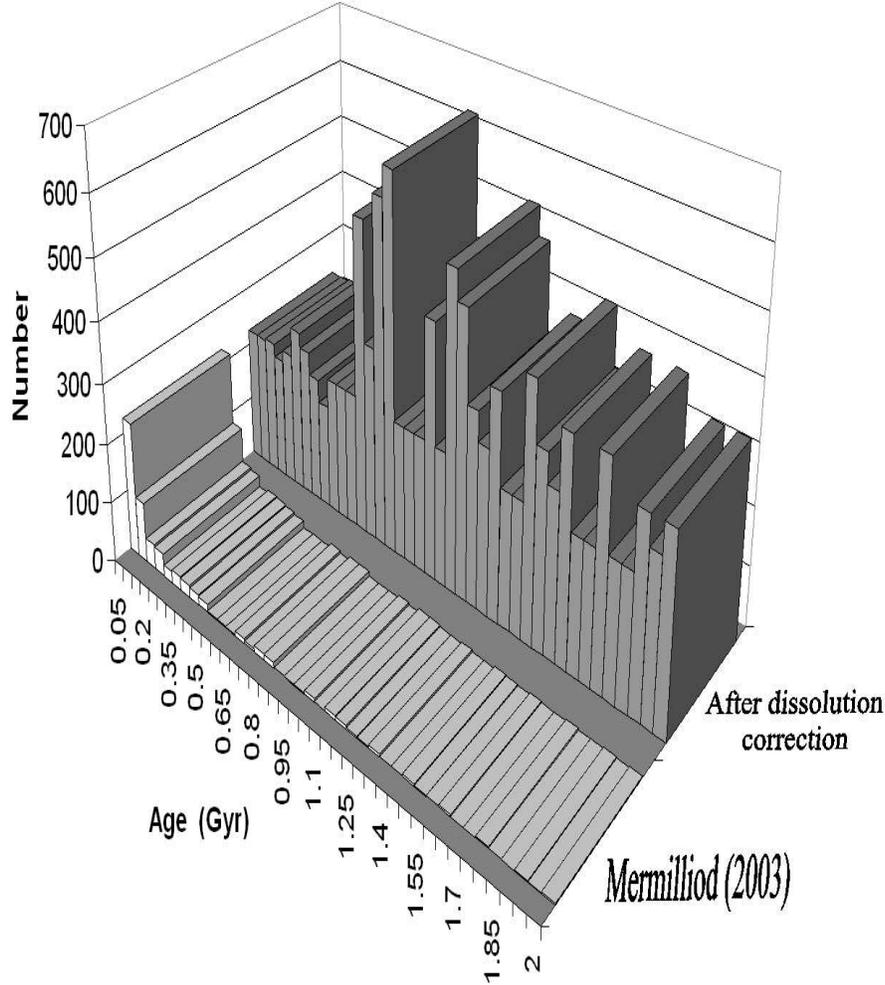,height=14cm,width=14cm,angle=0}
        }}
   \caption[]{Star formation history after the dissolution corrections
              (see the text), compared to the age histogram used 
              throughout this paper.}
      \label{best}
\end{figure*}
%
%
       \subsubsection{The 0.35 Gyr burst}
          This burst appears at 0.20-0.35 Gyr and it seems to be not
          very prominent.    

       \subsubsection{The 0.70 Gyr burst}
          This burst appears at 0.6-0.8 Gyr, lasting also 0.2 Gyr.
          It appears to be much stronger than burst 1. 

       \subsubsection{The 1.13 Gyr burst}
          It started 1.3 Gyr ago and lasted 0.35 Gyr. It was the
          strongest. The unusually high level of enhanced star
          formation could be the result of the combined action of
          density waves and a strong tidal interaction with the SMC.

       \subsubsection{The 1.50 Gyr burst}
          This burst appears at 1.4-1.6 Gyr, lasting also 0.2 Gyr.
          It appears to be similar in strength to burst 2.

       \subsubsection{The 1.93 Gyr burst}
          This burst started 2 Gyr ago, lasting 0.15 
          Gyr. Its activity was lower than for bursts 2 and 4 but much
          higher than 1. 

    \subsection{Volume correction: the solar neighbourhood}
       Since our original open cluster sample is not volume-limited,
       there could be a bias in the relative number of objects in each 
       age bin: open clusters with different initial populations and
       metallicities have different integrated luminosities, thus the
       volume of space sampled varies from cluster to cluster.
       In principle, our method will produce optimal results for 
       volume-limited open cluster samples. In this section we consider 
       a cluster sample in the solar neighbourhood. The solar neighbourhood 
       is defined as a volume centered on the Sun that is much smaller than 
       the overall size of the Milky Way galaxy and yet large enough to 
       contain a statistically useful sample of stars (see, e.g., Binney 
       \& Tremaine, 1987). The appropriate size of the volume depends on 
       which stars or objects are going to be investigated: for white 
       dwarfs, which are both common and faint, it may consist of a sphere 
       of radius 10 pc centred on the Sun, while for the bright but rare O 
       and B stars, the solar neighbourhood may be considered to extend as 
       far as 1 kpc from the Sun. In our analysis we will consider the 
       latter in order to retain a statistically significant sample of open 
       clusters. There are 161 open clusters in our sample within 1 kpc of 
       the Sun (and younger than 2 Gyr).
       
%
\begin{figure*}
        \centerline{\hbox{
        \psfig{figure=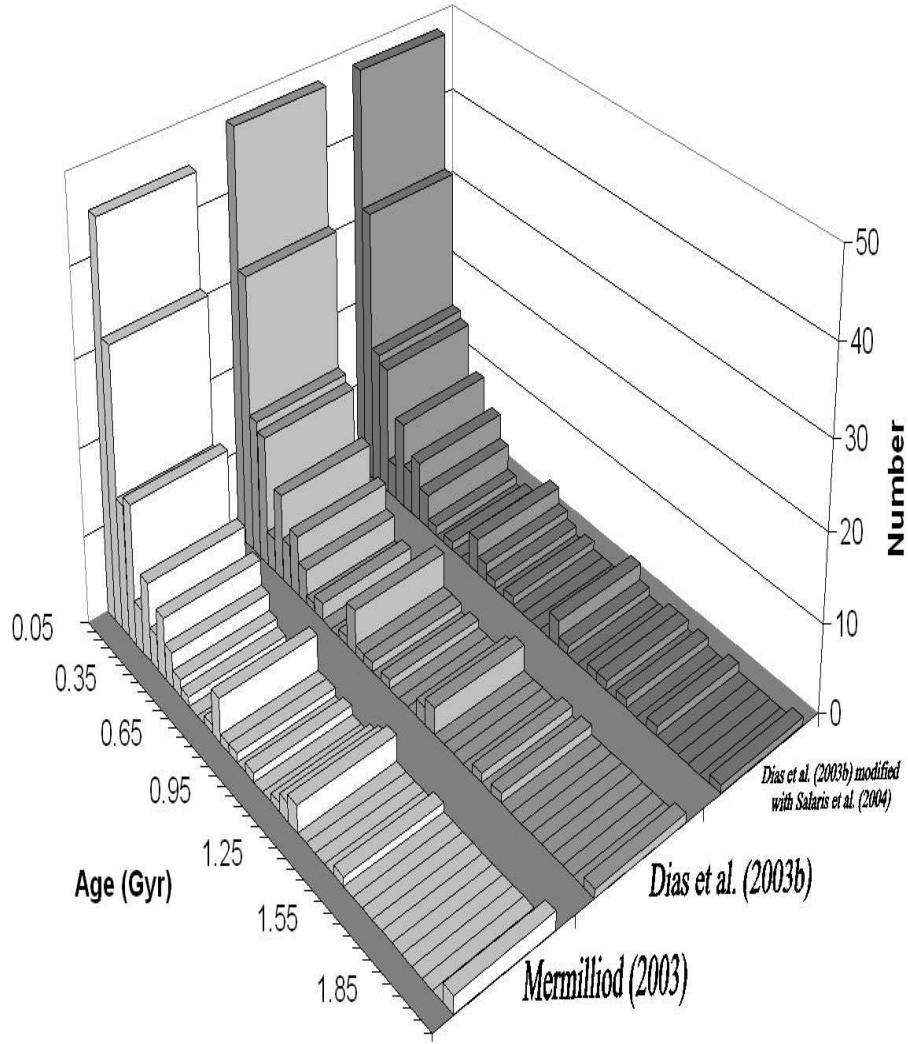,height=14cm,width=14cm,angle=0}
        }}
   \caption[]{Open cluster age distribution for objects in the three 
              samples within 1 kpc from the Sun, volume-limited sample. 
            }
      \label{sn}
\end{figure*}
%
%
       If we restrict our analysis to clusters closer than 1 kpc, the
       age distribution plotted in Fig. \ref{sn} is obtained. Burst 1
       is clearly identified at an age of 0.35 Gyr with a duration of 
       about 0.2 Gyr. Burst 2 is also present for an age of about 0.7
       Gyr with a duration of about 0.2 Gyr. Burst 3 appears at 1.15
       Gyr and is 0.1 Gyr wide. Unfortunately, it is difficult to recover
       burst 4 and 5 as the number of clusters older than 1.2 Gyr 
       within 1 kpc from the Sun is very low (3 objects). In fact, our
       results appear to suggest that the solar neighbourhood experienced
       quiescent star formation at 1.2-2 Gyr. From Fig. \ref{sn}, the
       star formation rate history in the solar neighbourhood seems to be
       slightly different, with not a well defined period and narrower 
       maxima. It is, however, easy to reconcile these apparent differences.
       The age distribution features interpreted as bursts 1, 2, and 3 are
       clearly identified in Fig. \ref{sn}. Burst4 and 5 may also exist
       but there are too few objects to confirm the hypothesis. On the other
       hand, uncertainties in the open cluster ages are smaller for closer
       clusters, therefore maxima are narrower and show better contrast
       against steady, quiescent star formation. 

\begin{figure*}
        \centerline{\hbox{
        \psfig{figure=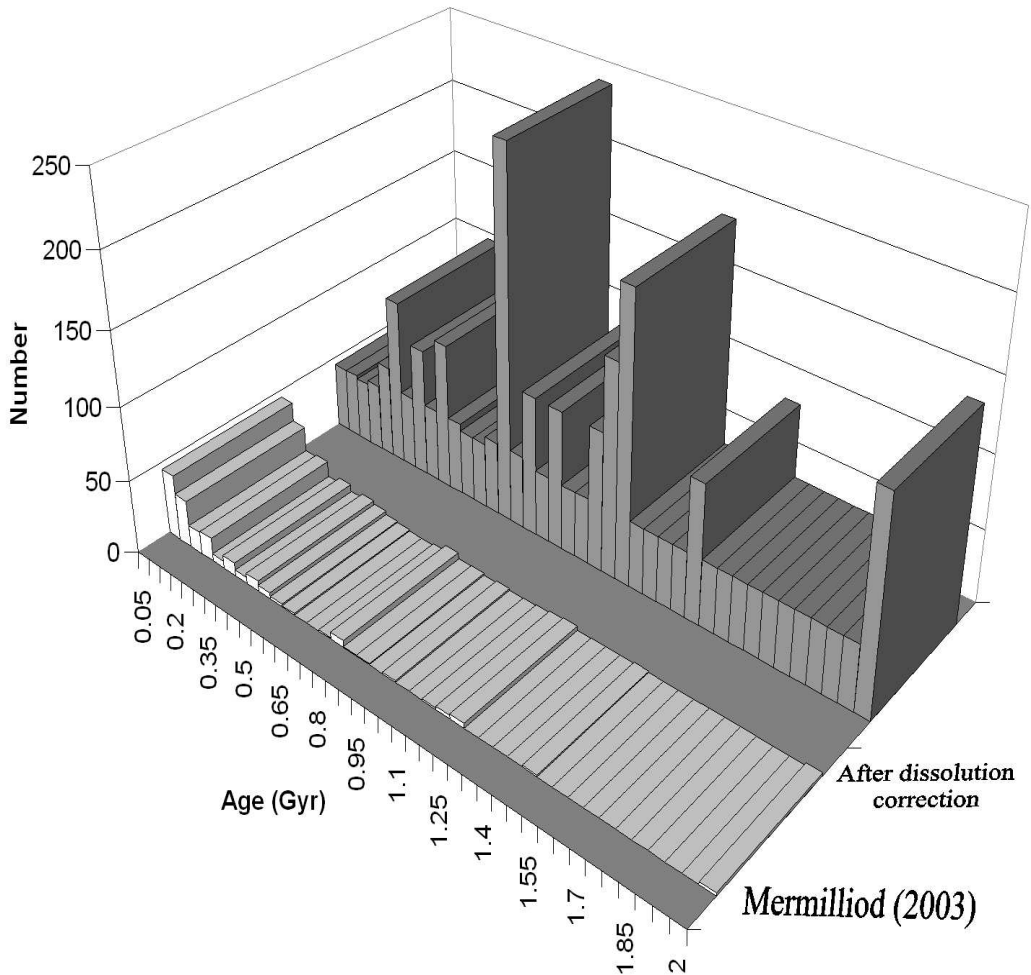,height=14cm,width=14cm,angle=0}
        }}
   \caption[]{Star formation history for the solar neighborhood 
              (volume correction) after the dissolution corrections
              (see the text), compared to the age histogram used 
              throughout this paper.}
      \label{bestp}
\end{figure*}

       Fig. \ref{bestp} shows the solar neighbourhood sample with 
       evolutionary correction as described in the previous section.
       In this case the second burst is the strongest, followed by
       the third, the fifth, and the first. The fourth one is now 
       the least important. 
       
%
%
%
%

 \section{Comparison with other authors}
    The problem of deducing the star formation history of our Galaxy
    has been primarily studied through chemical evolution models 
    although other methods have also been used: Hertzsprung-Russell 
    diagram inversion, stellar evolutionary tracks (Twarog, 1980; 
    Meusinger, 1991; Bertelli \& Nasi, 2001), chromospheric activity 
    as measured by Ca {\small II} and K emission (Barry, 1988; 
    Soderblom et al., 1991; Rocha-Pinto et al. 2000a, b), stellar 
    kinematics (Gomez et al., 1990; Marsakov et al. 1990; Chereul et 
    al., 1998; Just, 2002, 2003), the main-sequence luminosity function 
    (Scalo, 1987), the white dwarf luminosity function (Noh \& Scalo, 1990; 
    Diaz-Pinto et al., 1994; Isern et al., 1999), combining the 
    metallicity distribution and age-metallicity relation of G dwarfs
    (Rocha-Pinto \& Maciel, 1997) and the distribution of 
    coronal emission as measured by X-ray luminosities (Micela et al., 
    1993; Lachaume et al., 1999). Most of these studies have inferred an 
    star formation history that is non-monotonic with time. 

    Although the majority of studies in this field use samples of stars
    in the solar neighbourhood with no stars more distant than about
    100 pc being considered, it does not mean that the star formation
    history derived can only be applied to stars born in the solar
    neighbourhood. Nearby stars older than about 0.2 Gyr. come from 
    birth sites which span a large range in Galactocentric distances.
    Wielen (1977) showed that the orbital diffusion coefficient 
    deduced from the observed increase of velocity dispersion with age
    implies that such stars have suffered an rms azimuthal drift of
    about 2 kpc for an age of 0.2 Gyr. Considerable, but smaller, drift
    should occur also in the radial direction. Wielen et al. (1996),
    on the basis of the Sun's metallicity and the radial metallicity
    gradient in the Galactic disk, estimated that the Sun has migrated
    outward by 1.9$\pm$0.9 kpc in the past 4.5 Gyr. Sellwood and Binney
    (2002) have discussed the dynamics of radial migration due to transient
    spiral arms, and the relationship between radial migration and disk
    heating. They estimate that old stars formed in the solar neighbourhood
    should now be scattered nearly uniformly throughout an annulus of 
    8.5$\pm$4.0 kpc. In this sense the star formation history inferred for 
    nearby stars is a measure of the global star formation history, at least 
    at the Sun's Galactocentric radius. It is, however, not clear how the 
    diffusion process affects entire star clusters not standalone stars and 
    it is likely that star clusters experience little or no (azimuthal, 
    radial) drift over the life-time.     

    \subsection{Hernandez et al. (2000a)}
       Hernandez et al. (2000a) have derived the star formation history
       of the solar neighbourhood over the last 3 Gyr using data from 
       the {\it Hipparcos} catalogue to construct color-magnitude 
       diagrams. They treated these diagrams using advanced Bayesian 
       analysis techniques (Hernandez et al., 1999; Hernandez et al.,
       2000) to deduce the star 
       formation rate history of this region. They recovered the star 
       formation history with an unprecedented time resolution of 
       0.05 Gyr. This high resolution makes it easy to compare our 
       present results with those in their papers (Hernandez et al.
       2000a, b, 2001). Their results indicate that the local star 
       formation rate has an oscillatory component of period $\sim 0.5$ 
       Gyr superimposed on a small level of constant star formation 
       activity. Their inferred star formation history appears to be 
       compatible with the observed distribution of stars. This cyclic 
       behaviour is interpreted as the result of repeated encounters 
       with the Galactic arm density pattern.
       In their work, they find that the last episode of enhanced star
       formation happened about 0.5 Gyr ago. Another burst is found at
       about 1.3 Gyr and yet another one at 1.9 Gyr. Their Fig. 4 
       (Hernandez et al., 2000a) however suggests a double peak for the 
       youngest burst and also for the one at 1.3 Gyr. In fact, their
       results appear to indicate one burst at about 0.4 Gyr (our burst 
       1) and another one at 0.7 Gyr (our burst 2). The 1.3 Gyr burst 
       may be the convolution of burst 3 and 4 found in our analysis.
       Their third burst coincides in time with our fifth one. Therefore,
       our derivation of the recent star formation history both in the
       solar neighbourhood and the Galactic disk can be considered as
       fully consistent with the results found by Hernandez et al. (2000a)
       in spite of the fact that a completely different technique has been
       used in its derivation.

%
%
%
\begin{table}
   \caption[]{Main features of the star formation history compared with
              other authors.}
     \begin{center}
     \begin{tabular}{llll}
     \noalign{\smallskip}
     \hline 
          & This work & Hernandez et al. & Rocha-Pinto et al. 
            \\ 
          &           & (2000a)          & (2000a) 
            \\ 
     \noalign{\smallskip}
     \hline
     \noalign{\smallskip}
       Number of 'bursts'   &  5        &  3       & 1          \\
       Age of burst 1       &  0.4 Gyr &  0.5 Gyr & 0-1 Gyr    \\
       Age of burst 2       &  0.8 Gyr &  1.3 Gyr & -    \\
       Age of burst 3       &  1.2 Gyr &  1.9 Gyr & -    \\
       Age of burst 4       &  1.5 Gyr &  -       & -    \\
       Age of burst 5       &  1.9 Gyr &  -       & -    \\
     \noalign{\smallskip}
     \hline
     \end{tabular}
     \end{center}
     \label{comparison}
\end{table}
%
%
  
    \subsection{Other authors}
       Rocha-Pinto et al. (2000a) have derived a star formation history 
       from the chromospheric activity-age distribution of a large 
       sample comprising 552 stars with {\it Hipparcos} parallaxes in 
       the solar neighbourhood (stars more distant than 80 pc were 
       omitted), and have found intermittency in the star formation rate 
       over 14 Gyr. With a time resolution of 0.4 Gyr, their history of 
       the star formation rate plotted in Fig. 2 (Rocha-Pinto et al., 
       2000a) and Table 1 (Rocha-Pinto et al., 2000b) indicates that 
       the solar neighbourhood experienced a burst of star formation 
       0-1 Gyr ago (burst A). An epoch of low star formation took place 
       1-2 Gyr ago and a new, very strong burst was experienced 2-5 Gyr 
       ago (burst B). An additional burst was found 7-9.5 Gyr ago (burst
       C). Although their time resolution makes a direct comparison 
       impossible, their burst A is compatible with a convolution of our
       bursts 1 and 2, and possibly 3. There are, however, strong 
       discrepancies in the 1-2 Gyr range.
  
       Noh and Scalo (1990) using the white dwarf luminosity function found 
       peaks of star formation at 0.3 Gyr and 1.8 Gyr, that are compatible
       with our bursts 1 and 5.     
       Barry (1988) using a technique similar to the one in Rocha-Pinto
       et al. (2000a), has also found three bursts using a volume-limited
       sample of 115 F-G stars, the most recent peaked in the last 0.5 
       Gyr (burst A). An epoch of low star formation took place 
       1-4 Gyr ago and a new burst was experienced 4-6 Gyr ago (burst B). 
       The oldest burst was found 7-11 Gyr ago (burst C). His burst A appears
       to be a convolution of our bursts 1 and 2.
       Using the main-sequence luminosity function, Scalo (1987) found 
       signatures of starbursts 5-6 and 0.3 Gyr ago. His 0.3 Gyr burst
       coincides with our burst 1.
       Twarog (1980) found that during the last 4 Gyr the star formation
       rate has remained more or less constant but there was a sharp
       increase from 4 to 8 Gyr ago.
       Although the time resolution in these studies is rather low 
       (0.2-1.0 Gyr), their results seem to indicate that bursts 1, 2, and 5 
       have also been identified by other authors. 

 \section{Age distribution maxima: enhanced star formation vs.
          accretion origin}
    Starburst events are often seen in interacting galaxies, and it
    seems reasonable to infer a causal connection between galaxy
    interactions (mergers and close encounters) and starburst activity. 
    It is now becoming increasingly clear that average galaxies 
    engulf and devour smaller satellite galaxies as part of a 
    more general process of hierarchical merging. Our own galaxy, 
    the Milky Way, may have been incorporating dwarf galaxies into 
    its disk since it was formed. The Galaxy is encircled by
    satellite galaxies that appear confined to one of two great
    streams across the sky (Lynden-Bell \& Lynden-Bell, 1995). The
    most well-known of these are the Magellanic Clouds and the 
    associated HI Magellanic stream. All of these are expected to
    merge with the Milky Way in the distant future, largely due to 
    the dynamical friction from the extended halo. The first strong 
    evidence in favour of a hierarchical merging scenario for the 
    Milky Way was presented by Ibata et al. (1994, 1995, 1997) when 
    they discovered the Sagittarius dwarf galaxy, a low mass dwarf 
    spheroidal galaxy about 25 kpc from the Sun in an advanced state 
    of disruption that is being absorbed by the disk of the Milky Way. 
    In fact, our results appear to indicate that the Sagittarius Dwarf 
    Galaxy may have had some role in inducing enhanced star formation in 
    the Milky Way disk. For about 10 years, the Sagittarius Dwarf 
    Galaxy remained as the only strong evidence that the Milky Way is 
    incorporating satellite galaxies, but another object has been added 
    recently. The Canis Major Galaxy (Martin et al., 2004; Bellazzini et 
    al., 2004) appears to be a close relative of the Sagittarius Dwarf
    Galaxy in terms of structural parameters. This new finding seems to 
    confirm that part of the Galactic disk could be extragalactic in 
    origin. On the other hand, recent numerical models of disk galaxy 
    formation (Abadi et al., 2003a, b) indicate that disrupted galaxy 
    satellites might have contributed a significant fraction of the old 
    stars in the disk of the Milky Way. Navarro et al. (2004) have 
    confirmed that the Arcturus Group, a dynamically-coherent group of 
    10-12 Gyr old, metal-deficient stars located in the solar
    neighbourhood around the star Arcturus, are very likely of
    extragalactic origin. These authors suggest that the Group
    is part of the tidal debris of a 10$^8$ $M_{\odot}$ disrupted
    satellite accreted by the Milky Way about 8 Gyr ago.

    In this scenario, the peaks in the open cluster age distribution
    can be interpreted as signatures of merger events. This is
    however a non-straightforward interpretation as the variations 
    found in the star formation rate appear to be periodic in nature,
    and mergers are expected to occur randomly in time. However, the 
    actual sequence of events driving enhanced star formation episodes 
    could be even more complex, with both tidal interactions and 
    accretion events contributing to the global star formation history,
    as the unusual strength of burst 3 suggests. 
   
    For the main open cluster sample used in this work (Mermilliod, 2003),
    we find 97 clusters in the age range 0.2-0.6 Gyr and 80 objects in the
    age range 0.6-2.0 Gyr. There is a clear excess of clusters older than
    0.6 Gyr. Strong star bursts are able to explain this feature but it 
    could also be possible that some of these clusters may have been 
    incorporated by the Galactic disk during accretion events in the 
    recent past. In any case, the accretion scenario can only be important
    for coplanar interactions with the disk of the Milky Way and large
    accreted star clusters, otherwise the star clusters surviving the merger 
    would end up either destroyed or in the Galactic halo.

    \subsection{The Canis Major galaxy role}
       The chain of dynamical events that provoked the formation of
       the recently discovered Canis Major Dwarf Galaxy remnant 
       (Martin et al., 2003; Bellazzini et al., 2003) has been 
       proposed as the process that built up the Monoceros Ring
       (Newberg et al., 2002; Yanny et al., 2003; Ibata et al., 2003;
       Crane et al., 2003; Rocha-Pinto et al., 2003; Majewski et 
       al., 2004). An unusually high spatial density of open clusters 
       has been found in the
       field of the Canis Major galaxy. The list of open clusters
       includes, at least: NGC 2204, 2243, and 2477, Tombaugh 1 and 2, 
       Berkeley 20 and 33, Melotte 66, and AM-2. Bellazzini et al. 
       (2003) have also found that the position, distance, and stellar 
       population of the old open clusters AM-2 and Tombaugh 2
       strongly suggest that they are part of the Canis Major galaxy.
       Both are older than 4 Gyr and therefore they have not been included
       in our age distributions for clusters younger than 2 Gyr. 
       Nevertheless, if the physical association between these open 
       clusters and the merged dwarf galaxy is confirmed, it will be a 
       clear example of {\it intergalactic stellar pollution}, with 
       {\it foreign} star clusters being absorbed by the Galactic disk.
       If true, it may have a major impact on the determination
       of the star formation history of the Milky Way, with a fraction
       of the observed stars and open clusters being actually outsiders
       coming from other star formation histories.
    
    \subsection{Searching for dwarf galaxy remnants}
       In this paper we have shown that open clusters can be used as
       tracers to investigate the star formation history of the Milky 
       Way disk. If a number of dwarf galaxies with their cohorts of 
       star clusters have been captured and disrupted in the past, open 
       clusters absorbed from these galaxies can be used to find the 
       trail left by the parent galaxy. Searching for groups of coeval
       old open clusters with similar distances located in the same
       region of the sky can help to identify the bulk of the accreted 
       satellite.   
   
 \section{Discussion and conclusions} \label{dis}
    Does the Galaxy form stars continuously, or in bursts separated by
    epochs of relative quiescence? If star formation occurs in bursts,
    what processes mediate the bursts? These are just two of the many 
    questions that we have tried to answer in this paper.
    We have applied a new method to carry out an objective reconstruction
    of the star formation history of the Milky Way disk over the last 
    2 Gyr. A sample composed of 581 open clusters with known ages and
    distances was used in the derivation of this star formation history.
    Peaks in the age distribution diagram of the cluster sample were
    interpreted as signatures of star formation bursts. Our results 
    indicate that the analyzed star formation rate presents two 
    components: periodic episodes of enhanced star formation superimposed 
    on a quiescent star formation level. A constant star formation history 
    during the last 2 Gyr can therefore be discarded. A uniform rate of 
    cluster formation and an exponentially declining dissolution rate 
    would have produced a single population of clusters that would follow 
    a straight line in a logarithmic plot with a slope given by the 
    characteristic life-time of the population. Evidence for at least five
    epochs of enhanced star formation during the time interval studied
    has been found. The recent star formation history derived in this work is 
    consistent with star formation rate histories deduced using a range of 
    other techniques. Interpreting the age distribution diagram suggests
    that quiescent (non-enhanced) star formation has proceeded in discrete, 
    highly-correlated regions of activity producing clusters which 
    dissipate their parent molecular cloud on a time-scale of a few Myr 
    ($<$ 10 Myr). The majority of open clusters formed during the 
    non-enhanced star formation periods are destroyed and their members 
    become field stars on a time-scale $<$ 20 Myr. However, there 
    appears to be a long-lasting component as well, since some of the 
    clusters are able to survive for over 0.2 Gyr. During bursts of star 
    formation, the huge number of clusters formed make it possible to 
    produce larger than average objects that could survive for several Gyr.
    A number of simulations were done in order to estimate the impact of
    the age errors on the features found in the age distribution diagram.
    Even for age errors larger than 30\% some of these features (the 
    younger ones) are still clearly identifiable. On the other hand,
    we have examined the possibility that the Galactic bursts are coeval
    with features in the star formation history of the Magellanic Clouds
    and the Sagittarius Dwarf Galaxy, as well as with close encounters 
    between them and the Milky Way. Although the degree of uncertainty
    is large, there are several coincidences that suggest tidal 
    interactions can play some role as inductive forces of bursts of star 
    formation in the Milky Way disk. Distant galactic encounters may trigger 
    significant star formation but quiescent star formation in the disk 
    of the Milky Way appears to be characterized by a large number of 
    small, short lived open clusters that contribute a small fraction of 
    the total number of stars formed ($<$ 150 per cluster) with a small 
    number of large, relatively long lived clusters that may contribute 
    most of the stars in the disk.

    There is, however, a cyclic behaviour in the
    burst sequence that may be better explained by the density wave
    hypothesis (Lin \& Shu, 1964) for the presence of spiral arms in
    late-type galaxies. A model like the one outlined in 
    Hernandez et al. (2000a) and developed in Martos et al. (2004) can 
    explain easily the 0.4 Gyr periodicity that we detect.
    Heating by spiral structure can explain the main features of the 
    age-velocity dispersion relation and other parameters of the velocity
    distribution in the solar neighbourhood, like why the stars in a single
    velocity-space moving group have a wide range of ages (De Simone et al.,
    2004). For a pattern speed $\Omega_p$ = 20 km s$^{-1}$ kpc$^{-1}$
    (e.g. Martos et al., 2004) and a Galactocentric distance 
    $R_{\odot}$ = 8.5 kpc it implies an orbital period of about 1 Gyr
    for the Sun. If the enhanced star formation episodes are, in fact,
    due to the interactions with the spiral arms it means that our Galaxy has
    two arms. This has been recently suggested by Martos et al. (2004) using 
    a completely different approach.
    On the other hand, studies on the nature of star formation 
    triggering in large disks indicate that a number of mechanisms may 
    operate concurrently: gravitational instabilities, shocks between 
    colliding clouds of gas, enhanced pressure of the interstellar medium, 
    strong stellar winds, supernova shocks (star formation triggered by 
    nearby bursts of star formation), density waves, shear forces produced 
    by differential rotation, and interactions or mergers with other 
    galaxies. Star 
    formation triggered by previous star formation events is a
    self-propagating process which may continue over a much longer 
    period of time, hence it is possible that overall a relatively large 
    fraction of star formation is triggered.

    On the other hand, if the star formation history inferred from the
    age distribution of open clusters is consistent with that deduced
    for individual stars then one can consider this as an argument
    in favour of most of the observed stars being formed in some kind
    of star clusters. Since only a subsample of the young clusters are 
    likely to survive, an obvious question is whether most of the field
    stars in a galaxy are originally formed in clusters. Even in these
    young star forming regions many of the field stars are from clusters 
    that have already been dissolved, hence the true percentage of stars
    that were originally in clusters is even higher, and might conceivably
    be $\approx$ 100\%. It is tempting to suggest that perhaps the majority 
    of stars are formed in groups and clusters, and that the field stars are
    simply the remnants of the fainter, less dense clusters which have
    dissolved. In fact, infrared observations show that most star formation 
    occurs in embedded star clusters within Giant Molecular Clouds. These
    clusters have masses in the range 50-10$^{3}$ $M_{\odot}$ (Lada \& Lada, 
    2003). Nevertheless, it is still possible to consider the existence
    of two channels in the star formation process: the dispersed mode
    and the clustered mode, by modifying slightly our concepts of dispersed 
    and clustered. The dispersed mode would be represented by
    the numerous short lived ($\tau <$ 30 Myr) small star clusters that
    are part of the quiescent star formation rate. These are different
    from classical open clusters and may be called {\it Fast Living 
    Clusters} or {\it flusters} in opposition to the fully fledged 
    standard open clusters as the Hyades or Pleiades. The clustered mode
    would be represented by the relatively long lived open clusters.
    In our analysis of the last 0.2 Gyr of star formation,
    the number of clusters in the age range 30-100 remains relatively 
    constant as well as the fraction in the age range 100-200. This
    feature suggests a real gap in the mass distribution of disk star
    clusters: very few clusters are formed in the mass range 20-30 
    $M_{\odot}$ ($N$ = 50-75). To explain the very good matching between
    the star formation history inferred from individual stars and that
    from open star clusters we have to assume that although most of
    open clusters ($ >$ 90\%) are poor, and short lived ($N \le$ 200) most
    of the stars ($ >$ 90\%) are formed in larger clusters ($N >$ 200).
    This conclusion, however, appears not to be supported by current
    observations.

    The third largest galaxy in the Local Group, Triangulum galaxy (M33, 
    NGC 598) may be undergoing star formation events similar to the ones
    described in this paper. M33 is a spiral galaxy about half the size 
    of the Milky Way. Chandar et al. (1999) have found 44 young clusters 
    in M33 with ages $\leq$ 100 Myr and masses in the range $6 \ \times 
    \ 10^2$ to $2 \ \times \ 10^4 \ M_{\odot}$. Currently, M33 appears to 
    be forming many young compact clusters but less massive than an older 
    cluster population formed in a previous starburst with characteristic
    mass $\sim 10^5 \ M_{\odot}$. 

    Although our conclusions are uncertain as they are based on a relatively
    small (36\%) subsample of a larger but still incomplete sample composed 
    of almost 2000 open clusters, they are similar to the ones previously 
    found by a number of other independent studies. This 
    confirms the validity of our approach to infer star 
    formation histories (recent and old), not just for the Milky Way disk 
    but for any other galaxy. On the other hand, if the conclusions obtained
    in this paper are correct, the Galactic disk will start another episode
    of enhanced star formation within the next very few Myr if it is not 
    already happening right now.

 \begin{ack}
   We thank the referee whose comments greatly improved this paper.
   We are very grateful to S.J. Aarseth for providing his computer code
   and for his comments on the manuscript.  
   We thank the Department of Astrophysics of Universidad Complutense of 
   Madrid for providing excellent computing facilities at the Centro de 
   Proceso de Datos in Moncloa. The authors thank S. van den Bergh, L. O. 
   Lod\'en, J.-C. Mermilliod, G. Carraro, and I. Platais for very helpful 
   and interesting discussions. We also thank M. Martos, X. Hern\'andez 
   and associated group for sharing with us their results on the Galactic
   spiral pattern and its rotation speed in advance of publication. In 
   preparation of this paper we made use of the Open Cluster Database, 
   operated at the {\it Institut d'Astronomie de l'Universit\'e de 
   Lausanne}, Switzerland, the New Catalogue of Optically Visible Open 
   Clusters and Candidates, operated at the {\it Universidade de S\~ao 
   Paulo}, Brazil, the NASA Astrophysics Data System and the ASTRO-PH 
   e-print server.
 \end{ack}

\end{document}